# AGENCIES AND SCIENCE-EXPERIMENT RISK

*Eric E. Johnson*[*]

*There is a curious absence of legal constraints on U.S. government agencies undertaking potentially risky scientific research. Some of these activities may present a risk of killing millions or even destroying the planet. Current law leaves it to agencies to decide for themselves whether their activities fall within the bounds of acceptable risk. This Article explores to what extent and under what circumstances the law ought to allow private actions against such non-regulatory agency endeavors. Engaging with this issue is not only interesting in its own right, it allows us to test fundamental concepts of agency competence and the role of the courts.*

*Two case studies provide a foundation for discussion: NASA's use of plutonium power supplies on spacecraft, which critics say could cause millions of cancers in the event of atmospheric disintegration, and a Department of Energy particle-collider experiment that allegedly poses a small risk of collapsing the Earth. These extreme examples serve as a test-bed for applying insights from neoclassical economics, behavioral economics, risk-management studies, and cognitive psychology. The resulting analysis suggests that in low-probability/high-harm scenarios, agencies are likely to do a poor job of judging the acceptability of risk to the public. Instead, generalist judges working in a common-law vein may have surprising advantages. This in turns suggests that under certain circumstances the government should be subject to legal action that provides non-*

* Associate Professor of Law, University of North Dakota School of Law. For helpful feedback and productive conversations, I thank Michael Baram, Rhett Larson, Kali Murray, Dmitry Bam, Priya Gupta, Monu Bedi, Meredith Johnson Harbach, Tara Helfman, Matt Parlow, Chad Oldfather, Olivier Moréteau, John Devlin, Ken Levy, Michael Coenen, Edward P. Richards, Margaret S. Thomas, Jack M. Weiss, Katheleen Guzman, Eric C. Chaffee, Browne Lewis, James C. Smith, Sarah Burstein, Monika Erhman, Megan Shaner, Stephen Galoob, Gus Hurwitz, Jill Wieber Lens, Melissa Luttrell, and Kit Johnson. For research assistance, I thank Jan Stone and Anne Mostad-Jensen. I also wish to thank the organizers of and participants in the Second Annual Junior Faculty Works-in-Progress Conference at Marquette University Law School and the 2014 Central States Law School Association Annual Conference at Louisiana State University.

*deferential review of discretionary agency actions that are non-regulatory in nature.*

TABLE OF CONTENTS

## I. INTRODUCTION

When getting ready to test the first atomic bomb, scientists of the Manhattan Project considered the possibility that detonating the device might ignite a runaway chain reaction in the atmosphere, engulfing the world in a fireball that would kill all plant and animal life.[1] They wagered it would not and threw the switch.

That bit of history is fairly well known. What is not widely appreciated, however, is that government agencies make similar decisions quite often—deciding to take small chances of causing enormous catastrophe. This Article considers how the law ought to constrain agencies in exposing the public to catastrophic risk created by agencies' own conduct.

---

1. *See* H.A. Bethe, *Ultimate Catastrophe?*, 32 BULL. ATOMIC SCIENTISTS, June 1976, at 36, 36.

There can be a stark difference in what risks administrative agencies tolerate when it comes to their own conduct versus the conduct of private actors. In regulating private actors, for instance, the Environmental Protection Agency ("EPA") and the Food and Drug Administration ("FDA") frequently use a one-in-a-million chance of killing a single person as a trigger for agency action.[2]

Contrast that with the decision-making of the Department of Energy ("DOE") about its own nuclear experiments. In 2000, a government lab started up a particle accelerator—the Relativistic Heavy Ion Collider—after certain risk modeling indicated a not-greater-than one-in-10,000 risk that the experiment might create a dangerous particle of "strange matter."[3] After a latency period of years or decades, it is hypothesized, such an object could initiate a runaway process that physically destroys Earth, extinguishing all life.[4] This risk model left "a comfortable margin of error" according to the report commissioned by the lab.[5] And this risk question is not a historical relic. Today, the RHIC has been upgraded and its program extended, with no new safety assessments having been done in the interim.[6]

The DOE, moreover, is not alone in generating exotic questions of science-experiment risk. Federal agencies rearrange the genetics of deadly viruses, operate experimental nuclear reactors, and launch rockets car-rying radioactive materials.[7] According to critics, some of these activities might endanger millions of lives.[8]

---

2. *See infra* note 135 and accompanying text.

3. *See* Adrian Kent, *A Critical Look at Risk Assessments for Global Catastrophes*, ARXIV:HEP-PH/0009204v7, July 24, 2015, at 3–5, http://arxiv.org/pdf/hep-ph/0009204v7.pdf [hereinafter Kent, *Critical Look*]. See *infra* note 134 and accompanying text for a discussion of this figure. Note that report authors released a second version of their report where they increased this model's probability bound to one in 100,000. *See infra* note 137 and accompanying text.

4. *See* Kent, *Critical Look*, *supra* note 3, at 3–4, 6–7.

5. *See id.* at 5; W. BUSZA, R.L. JAFFE, J. SANDWEISS, & F. WILCZEK, REVIEW OF SPECULATIVE "DISASTER SCENARIOS" AT RHIC at 23 (1999), http://www.bnl.gov/rhic/docs/rhicreport.pdf [hereinafter BUSZA ET AL.].

6. *See, e.g.*, *Black Holes at RHIC?*, BROOKHAVEN NAT'L LABORATORY, https://www.bnl.gov/rhic/blackHoles.asp (last visited Jan. 24, 2016) [hereinafter *RHIC Black Holes?*] (referring to 1999 safety report).

In *Palsgraf v. Long Island Railroad,* Judge Andrews observed, "Every one owes to the world at large the duty of refraining from those acts that may unreasonably threaten the safety of others."[9] If that holds true in the *Palsgraf* context of exploding fireworks at a commuter rail station, then one imagines governments must owe a duty to refrain from unreasonably threatening destruction on a global scale. Yet questions of low-probability/high-harm risk posed by agencies' own activities have generally evaded judicial review.

How can this be? At the threshold, the doctrine of sovereign immunity prevents most suits against federal agencies absent a waiver of immunity. Such waivers exist, but they may be near-misses when it comes to challenging potentially catastrophic government laboratory actions.

The Federal Tort Claims Act ("FTCA")[10] constitutes a waiver of sovereign immunity, but, for several reasons, it cannot provide a judicial-review mechanism for agency-created science-experiment risk. The FTCA's discretionary-function exception prevents lawsuits that challenge agency decisions as to whether their experimental programs pose an unacceptable risk.[11] In addition, the FTCA does not allow for injunctions, instead permitting only an award of compensatory damages after damage has already been done.[12] And in low-risk/high-harm scenarios such as those considered here, the theoretical possibility of after-the-fact damages is vanishingly unlikely to affect ex ante behavior.

---

7. *See, e.g.*, Michael Baram, *Biotechnological Research on the Most Dangerous Pathogens: Challenges for Risk Governance and Safety Management*, 47 SAFETY SCI. 890, 894–95 (2008) (discussing various incidents, including an incident at CDC, at a highest-level biosafety lab where a combination of lightning and damage to a cable from construction "shut down the negative air pressure system needed to keep dangerous select agents from escaping containment"); Alison Young & Nick Penzenstadler, *Inside America's Secretive Biolabs: Investigation Reveals Hundreds of Accidents, Safety Violations and Near Misses Put People at Risk*, USA TODAY (May 28, 2015), http://www.usatoday.com/story/news/2015/05/28/biolabs-pathogens-location-incidents/26587505/ (noting that the Federal Select Agent Program "inspects and regulates the subset of research labs that experiment with about four dozen types of pathogens deemed to pose bioterror threats"); Suzanne Baker, *Accident Tolerant Fuels Ready for Testing,* PHYS.ORG (Feb. 10, 2015), http://phys.org/news/2015-02-accident-tolerant-fuels-ready.html (describing testing of nuclear fuels "with enhanced tolerance to severe accident conditions" in Idaho National Laboratory's Advanced Test Reactor); *see infra* Part II.B–C (discussing particle accelerators and rockets with radioactive materials).

8. *See, e.g.*, Young & Penzenstadler, *supra* note 7 (providing that the "consequences could be devastating if accidents were to occur with lab-created strains of deadly influenza viruses that are purposely engineered to be easier to spread than what's found in nature").

9. 248 N.Y. 339, 350 (1928) (Andrews, J., dissenting).

10. Federal Tort Claims Act, ch. 753, title IV, 60 Stat. 812, 842-47 (1946) (codified as amended at 28 U.S.C. §§ 1346(b), 2671–2680 (2012)).

11. 28 U.S.C. § 1346(b)(1) (allowing claims against the United States "for money damages . . . for injury or loss of property, or personal injury or death caused by the negligent or wrongful act . . . of any employee of the Government . . . under circumstances where the United States, if a private person, would be liable to the claimant"); *id.* § 2680(a) (providing that the Act shall not apply to claims "based upon the exercise or performance or the failure to exercise or perform a discretionary function or duty on the part of a federal agency or an employee of the Government, whether or not the discretion involved be abused").

12. *Id.* § 1346(b)(1).

Federal law does have a system specifically designed for reigning in agency discretion—the Administrative Procedure Act ("APA").[13] Yet while the APA provides a rich system of checks and balances for agency rulemaking and adjudication, it has almost nothing to say about private-actor-type agency action that plausibly threatens public safety. While catch-all provisions of the APA are capable of supporting judicial review in such circumstances, the statute's lack of an explicit invitation makes it all too easy for courts to avoid undertaking the task.

Theoretically, all government action is subject to direct constitutional challenge. Laboring under self-imposed doctrines of judicial restraint, however, the courts are loath to go where neither the APA nor FTCA has explicitly invited them. That reluctance is understandable given the received wisdom about the two statutes. It is widely understood that the APA "was designed to govern both internal agency procedure and judicial review and [has been] thought to be complete enough to cover the whole field."[14] And the courts have recognized that with the FTCA's discretionary-function exception, "Congress wished to prevent judicial 'second-guessing' of legislative and administrative decisions grounded in social, economic, and political policy through the medium of an action in tort."[15]

This is why potential government lab disasters are so interesting as a legal matter: They illuminate an important gap between the APA and FTCA. And it is clear that this void is not by design. The APA and FTCA, both passed in June 1946, are creatures of the era before big agency science.[16] Accordingly, they never comprehended today's exotic agency hazards. Similarly, judicial self-restraint doctrines rest on assumptions that predate, and so take no account of, current risks.

The legal gap might be less troubling if it were not for insights from behavioral economics, neoclassical economics, cognitive psychology, and the risk-management literature, all of which indicate that agency scientists are prone to misjudging how risky their activities really are.[17] Moreover, political control of agencies is inadequate when it comes to the pro-

13. Administrative Procedure Act, Pub. L. No. 79-404, 60 Stat. 237 (1946) (codified as amended at 5 U.S.C. §§ 501–706 (2012)).

14. John F. Duffy, *Administrative Common Law in Judicial Review*, 77 TEX. L. REV. 113, 114–15 (1998) (internal quotation marks omitted) (citing S. Rep. No. 79-752, at 7 (1945)).

15. United States v. S.A. Empresa de Viacao Aerea Rio Grandense (Varig Airlines), 467 U.S. 797, 814 (1984).

16. In 1946, at the time of the passage of the APA and FTCA, the atmospheric-ignition issue from the 1945 atomic bomb test had not been made public. *See* Bethe, *supra* note 1, at 36–37.

17. *See infra* Part IV.

spect of unprecedented laboratory catastrophes. This is in no small part because it is hard to take seriously what sounds like the stuff of science fiction.

This Article looks at two case studies. First is NASA's launch of plutonium power modules, one of which could, if it were to burn up on re-entry, disperse enough radioactive material to kill thousands and perhaps millions of people.[18] Second is the DOE's heavy-ion-collision experiments, referenced above, which critics say might initiate a process that physically crushes the Earth into an ultradense ball of strange matter.[19] No serious commentators suggest the probabilities of such events are anything but small. Yet there are two reasons we should pay attention anyway.

First, where the potential harm is very large, even tiny probabilities of disaster may be statistically significant. In the mathematics of risk, a one-in-10,000 risk of a million deaths can be equated to 100 certain deaths. And a one-in-10,000 risk of the loss of the entire human population would equate to more than 700,000 certain deaths.

Second, even though the probability of catastrophe might be small for any given research program, the trend is toward federal agencies doing more and more large-scale research. Eventually the odds of disaster may catch up with us. Thus, it makes sense now to ensure that our legal institutions will engage with such risk questions in a way that is meaningful, rational, and effective.

So far, this subject has fallen between the cracks. In the administrative procedure literature, commentators have explored in detail the deference given government agencies in rulemaking and adjudication.[20] But agency discretion with regard to non-rulemaking/non-adjudicatory activity is largely unaddressed. In the risk literature, scholars have written a great deal about what role administrative agencies should play in addressing risks of catastrophic disasters posed by private organizations, particularly those exploiting nuclear, chemical, and biological technolo-gies.[21] But scholars have paid little attention to catastrophic risks that agencies themselves pose through their own laboratory research activities. By putting a spotlight on catastrophe risk created by direct agency action, this Article seeks to contribute to both discussions.

---

18. *See infra* Part II.B.

19. *See infra* Part II.C.

20. A multitude of scholarship has been written, for instance, on judicial deference to agency interpretations of statutory authority to engage in rulemaking and the role of *Chevron U.S.A., Inc. v. Natural Res. Defense Counsel, Inc.*, 467 U.S. 837 (1984). *See, e.g.*, Thomas J. Miles & Cass R. Sunstein, *Do Judges Make Regulatory Policy? An Empirical Investigation of* Chevron, 73 U. CHI. L. REV. 823 (2006) (demonstrating relationship between deference to agency interpretations and ideology of judges); Nina A. Mendelson, Chevron *and Preemption*, 102 MICH. L. REV. 737 (2004) (discussing deference to agency statutory interpretation as an impingement on federalism); Thomas W. Merrill & Kristin E. Hickman, Chevron*'s Domain*, 89 GEO. L.J. 833 (2001) (analyzing which agency interpretations should be given mandatory deference); Jonathan T. Molot, *The Judicial Perspective in the Administrative State: Reconciling Modern Doctrines of Deference with the Judiciary's Structural Role*, 53 STAN. L. REV. 1 (2000) (positing a beneficent judicial influence on the legislature in favor of fairness and rationality that is undermined by deference to agency statutory interpretation).

Part II of this Article explores the particular kinds of risks at issue here: irreversible and catastrophic ultrahazards. This Part includes discussion of this Article's case studies—NASA's plutonium-fueled spacecraft and the DOE's heavy-ion collider. Part III surveys the legal landscape of administrative procedure and judicial doctrine that shields agency decisions about acceptable risk from legal oversight and court challenge. Part IV explains why the unreviewability of agency science experimentation is problematic, looking at a number of insights from the social sciences that provide reason to believe agencies will not do an adequate job of protecting the public from hazards that agencies themselves create. Part V discusses the solution—courts entertaining suits against federal agencies plausibly posing irreversible and catastrophic risks through their non-regulatory/non-adjudicatory actions.

## II. AGENCY SCIENCE EXPERIMENTS AND PECULIAR RISKS

There are innumerable risks in the world, but this Article concentrates on a particular species of uniquely troublesome risks—what we can call *catastrophic and irreversible ultrahazards*. After discussing some general features of this category of risk, I will look at two examples: NASA's plutonium-powered spacecraft and the DOE's heavy-ion collider program.

### *A. Catastrophic and Irreversible Ultrahazards*

The risks I am discussing are, in a technical sense, "ultrahazards." That is, they are the kinds of risks for which the common law deems strict liability appropriate. These are risks that arise where, as Richard A. Epstein puts it, "small triggers . . . can release far larger forces."[22] In *Rylands*

---

21. Regarding nuclear or radiological risk, see, e.g., Albert C. Lin, *Lessons from the Past for Assessing Energy Technologies for the Future*, 61 UCLA L. REV. 1814 (2014); Joseph P. Tomain & Constance Dowd Burton, *Nuclear Transition: From Three Mile Island to Chernobyl*, 28 WM. & MARY L. REV. 363 (1987); Diane Carter Maleson, *The Historical Roots of the Legal System's Response to Nuclear Power*, 55 S. CAL. L. REV. 597 (1982). Regarding chemical risk, see, e.g., Lauren Trevisan, *Human Health and the Environment Can't Wait for Reform: Current Opportunities for the Federal Government and States to Address Chemical Risks Under the Toxic Substances Control Act*, 61 AM. U. L. REV. 385, 385 (2011); Noah M. Sachs, *Jumping the Pond: Transnational Law and the Future of Chemical Regulation*, 62 VAND. L. REV. 1817, 1818 (2009). Regarding biological risk, see, e.g., Barry Kellman, *Regulation of Biological Research in the Terrorism Era*, 13 HEALTH MATRIX 159 (2003); John Miller, Note: *Beyond Biotechnology: FDA Regulation of Nanomedicine*, 4 COLUM. SCI. & TECH. L. REV. 1 (2003); Barry Kellman, *Biological Terrorism: Legal Measures for Preventing Catastrophe*, 24 HARV. J.L. & PUB. POLICY 417, 419 (2001).

22. RICHARD A. EPSTEIN, TORTS 348 (1999)

*v. Fletcher,* the case credited with creating the modern conception of strict liability, the court held that whoever "brings, or accumulates, on his land anything which, if it should escape, may cause damage to his neighbor . . . does so at his peril."[23] In sum, ultrahazards combine small triggers, large forces, and a threat to neighbors.

In this Article, however, I am not interested in ultrahazards generally. My focus is on that subset of risks that are catastrophic and irreversible. In *Rylands*, because of the availability of compensatory damages, the defendant created an ultrahazardous condition *at his own peril.* By contrast, catastrophic and irreversible risks are *at everyone's peril.* When it comes to killing millions of people or permanently transforming the planet, an award of compensatory damages cannot make the defendant shoulder the burden.

Because of inadequacy of after-the-fact damages, incipient catastrophic and irreversible harms are appropriate subjects for before-the-fact legal restrictions. Ex ante prohibitions familiarly come in two forms—court-issued injunctions and agency-issued regulations. Given these established forms of legal constraint, catastrophic and irreversible ultrahazards might seem easy to deal with as a legal matter. Who would not agree that it is prudent to halt an activity in which a small trigger threatens to unleash great destruction? But there is a wrinkle: These risks routinely involve low or uncertain probabilities, and the harmful effects commonly involve some degree of latency. The combination of latency and long odds makes these risks easy to ignore—even if doing so is unwise.[24]

To further pin down the sorts of risk I am looking at here, it makes sense to put them in the context of the "precautionary principle" debate. The precautionary principle is the idea that it makes sense to take precautionary measures against plausible threats even in the absence of a full empirical demonstration of a cause-and-effect relationship between an activity and the alleged harm.[25] The debate over the precautionary principle in the legal scholarly literature points up how catastrophic and irreversible risks are in a unique category.[26] Cass R. Sunstein, a leading critic in the debate, has assailed the precautionary principle's general "incoherence."[27] Yet when it comes to catastrophic and irreversible risks, even Sunstein has pronounced the precautionary principle "sensible."[28] I should emphasize that in looking at catastrophic and irreversible *ultra-hazards*, I am talking about an even smaller subset of risk than those to which Sunstein would apply the precautionary principle. For example, Sunstein would apply the precautionary principle to climate change—since it is catastrophic and irreversible. But climate change does not qualify as an ultrahazard because it does not have a small trigger. To the contrary, climate change has a very large trigger—it takes whole societies working over the course of decades to put enough carbon dioxide into the atmosphere to change the climate. The kinds of disasters I am talking about have very small triggers—for instance, in our two case studies, a single error in navigational software or a single collision of nuclei in a particle accelerator.[29]

---

23. [1868] 3 L.R.E. & I. App. 330, 340 (appeal taken from Eng.).

24. *See, e.g.*, Paul Slovic, *Perception of Risk*, 236 SCIENCE 280, 280 (1987) ("In recent decades, the profound development of chemical and nuclear technologies has been accompanied by the potential to cause catastrophic and long-lasting damage to the earth and the life forms that inhabit it. The mechanisms underlying these complex technologies are unfamiliar and incomprehensible to most citizens. Their most harmful consequences are rare and often delayed, hence difficult to assess by statistical analysis and not well suited to management by trial-and-error learning."); *see also* Clayton P. Gillette & James E. Krier, *Risk, Courts, and Agencies*, 138 U. PA. L. REV. 1027, 1077 (1990) ("Latency and irreversibility practically deny us the fruits of trial-and-error, perhaps the best means yet devised by which to resolve uncertainty.").

25. Douglas A. Kysar, *It Might Have Been: Risk, Precaution and Opportunity Costs*, 22 J. LAND USE & ENVTL. L. 1, 4 (2006).

In sum, the exotic agency-science risks discussed here constitute a truly elite set of menaces. That makes it all the more remarkable that our legal structure refrains from engaging with them. In the following two subparts, I explore two real-life examples of alleged catastrophic and irreversible ultrahazards. Together, they will help to flesh out the unique features of these risks and provide a model for the analysis that follows in Parts III and IV.

### *B. NASA and Atmospheric Dispersal of Plutonium-238*

NASA's Jet Propulsion Laboratory ("JPL") sends robotic spacecraft to explore the solar system.[30] Their projects have included the Voyager probes that captured iconic images of Jupiter and Saturn, as well as the Mars rovers that fueled the imagination of many an elementary school science class.[31] All robotic exploration spacecraft need power for their on-board electronics. Some use solar panels, taking their energy from sunlight. Others get their electricity from radioisotope thermoelectric generators ("RTGs").[32] An RTG uses heat to generate electricity,

---

26. For a sample of the debate, see Cass R. Sunstein, *Irreversible and Catastrophic: Global Warming, Terrorism, and Other Problems*, 23 PACE ENVTL. L. REV. 3, 10 (2006) [hereinafter Sunstein, *Irreversible and Catastrophic*]; David A. Dana, *A Behavioral Economic Defense of the Precautionary Principle*, 97 NW. U. L. REV. 1315, 1320–21 (2003) (arguing that the precautionary principle is justifiable because of behavioral economic effects that tend toward underweighting of risks).

27. Sunstein, *Irreversible and Catastrophic*, *supra* note 26, at 15.

28. *Id.* at 5.

29. *See infra* Part II.B–C.

30. *See* NASA, NASA FACTS: JET PROPULSION LABORATORY 1 (2011), http://www.jpl.nasa.gov/news/fact_sheets/jpl.pdf.

31. *See id.* at 2–3.

32. Radioisotope thermoelectric generators are, in particular, used for spacecraft going to the outer solar system where sunlight is feeble. *See* NASA, NASA FACTS: RADIOISOTOPE POWER and that heat is generated by a mass of highly radioactive material.[33] The plutonium in an RTG is so radioactive that it is literally hot—so hot, in fact, that it glows bright orange like the heating element of an electric stove.[34]

Rockets and spacecraft, of course, occasionally go awry. Thus, critics have alleged that NASA's launches of RTG-powered spacecraft pose a public health risk. The worry is that a vehicle failure could release plutonium into the environment where it could cause cancer and other health problems.

Understanding the risk issue with RTGs requires understanding how they are used, how they work, and what sort of radiological health threat they are said to represent. So I'll provide a brief overview.

At the outset, it is important to observe that RTGs do not work like nuclear reactors.[35] There is no nuclear fission chain reaction. Thus, a mass of plutonium-238 on a spacecraft is not able to excite itself into a meltdown or a nuclear-weapon-type explosion.[36] An atom of plutonium with one more neutron is plutonium-239. That's the fissionable isotope used in nuclear bombs—what's often called "weapons grade" plutonium.[37] Plutonium-238, is, as a general matter, merely radioactive. Its atoms decay, releasing heat and alpha particles. What's more, alpha particles are

---

SYSTEMS: MISSION NEED 2 (2012), https://solarsystem.nasa.gov/rps/docs/APPRPS%20MissionNeed FactSheet9-27-12.pdf. Spacecraft staying aloft in the inner solar system generally use solar panels. *See, e.g.*, *Electrical Power*, NASA, http://mars.nasa.gov/mro/mission/spacecraft/parts/electricalpower/ (last visited Jan. 24, 2016). For other missions—Mars landers, for instance—NASA has sometimes chosen solar power, and sometimes RTGs. *Compare* NASA, NASA FACTS: MARS EXPLORATION ROVER 2 (2004), http://www.jpl.nasa.gov/news/fact_sheets/mars03rovers.pdf (solar panels on rovers Spirit and Opportunity), *with* NASA, NASA FACTS: MARS SCIENCE LABORATORY/CURIOSITY 3 (2013), http://www.jpl.nasa.gov/news/fact_sheets/mars-science-laboratory.pdf (plutonium-238 RTG used on rover Curiosity). For general background on RTGs, see *Radioisotope Thermoelectric Generator (RTG)*, NASA, https://solarsystem.nasa.gov/rps/rtg.cfm (last visited Jan. 24, 2016).

33. *See* NASA, NASA FACTS: RADIOISOTOPE POWER SYSTEMS FOR SPACE EXPLORATION 1–2 (2011), https://solarsystem.nasa.gov/docs/APP%20RPS%20FactSheet%203-31-11.pdf.

34. *See Radioisotope Power Systems: Safety*, NASA, https://solarsystem.nasa.gov/rps/safety.cfm (last visited Jan. 24, 2016).

35. For background on how RTGs work, see *Radioisotope Thermoelectric Generator (RTG)*, *supra* note 32; NASA, NASA FACTS: WHAT IS PLUTONIUM-238? 1 (2012), https://solarsystem.nasa.gov/rps/docs/APP%20RPS%20Pu-238%20FS%2012-10-12.pdf; Dave Mosher, *NASA's Plutonium Problem Could End Deep-Space Exploration*, WIRED (Sept. 19, 2013, 6:30 AM), http://www.wired.com/2013/09/plutonium-238-problem/.

36. *See generally id.* WHAT IS PLUTONIUM-238?, *supra* note 35, at 1, notes, "Pu-238 would not work well as the fuel in a nuclear reactor and it is not the type of plutonium used for nuclear weapons."

37. Plutonium-239 is useful for nuclear weapons and reactors because of its ability to chain react. A Plutonium-239 atom can spontaneously undergo nuclear fission, a process by which it breaks into two or more smaller nuclei and at the same time releases free neutrons a large amount of energy. The neutrons released by one nucleus can then induce fission on contact with neighboring nuclei, causing a chain reaction. *See generally Physical, Nuclear, and Chemical Properties of Plutonium*, INST. FOR ENERGY & ENVTL. RES., http://ieer.org/resource/nuclear-power/plutonium-factsheet/ (last visited Jan. 24, 2016).

the weaklings of the radioactivity world—they can be stopped by thin clothing or even a few inches of air.[38]

But plutonium-238 is not harmless. The isotope poses a severe health problem if it is inhaled or ingested.[39] In that case, there is no air or clothing between the plutonium and the body's cells, and, thus, there is nothing to stop the alpha particles from energetically piercing cellular membranes and precipitating DNA mutations—mutations that can transform a normal cell into a cancer cell.[40] An accident involving an RTG, therefore, has the potential to work like a radiological weapon or "dirty bomb"—spreading radioactive contamination over a wide area.[41]

Concern over RTG safety boiled over in 1997 in the months leading up to the launch of the Cassini probe to Saturn. Loaded with 72.3 pounds of plutonium-238,[42] Cassini inspired a "Stop Cassini" protest movement.[43] NASA responded to the criticism by emphasizing the safety measures it had taken. While NASA acknowledged that plutonium-238 could cause cancer, the agency said that it mitigated the risk by sealing the plutonium into hardened pellets that would survive an explosion, a fire, or a barrage of shrapnel.[44] Moreover, NASA said the risk of a launch accident was low—ranging from between one-in-476 to one-in-1,100, depending on the

38. *See, e.g.*, WHAT IS PLUTONIUM-238?, *supra* note 35, at 1 ("Alpha particles emitted by Pu-238 decay can be blocked by simple barriers such as a thin sheet of paper.").

39. *See id.* at 2; *Plutonium – PU*, LENNTECH, http://www.lenntech.com/periodic/elements/pu.htm (last visited Jan. 24, 2016).

40. Plutonium's health hazards are discussed in U.S. NUCLEAR REGULATORY COMMISSION, OFFICE OF PUBLIC AFFAIRS, BACKGROUNDER: PLUTONIUM 1 (May 2014) (on file with author), http://www.nrc.gov/reading-rm/doc-collections/fact-sheets/plutonium.pdf ("Plutonium predominantly emits alpha particles—a type of radiation that does not penetrate and has a short range so it is easy to contain. But plutonium can be long-lived. It can deposit in the bones and lungs, and could increase an individual's cancer risk. In spite of its potential dangers, plutonium has several unique properties that make it useful."). That government factsheet, however, has since been replaced with one, dated September 2015, that no longer mentions "cancer" or "dangers." *See* U.S. NUCLEAR REGULATORY COMMISSION, OFFICE OF PUBLIC AFFAIRS, BACKGROUNDER: PLUTONIUM 1 (Sept. 2015), http://www.nrc.gov/reading-rm/doc-collections/fact-sheets/plutonium.html ("Plutonium predominantly emits alpha particles–a type of radiation that is easily stopped and has a short range. It also emits neutrons, beta particles and gamma rays. It is considered toxic, in part, because if it were to be inhaled it could deposit in lungs and eventually cause damage.").

41. *See, e.g*, JOHN R. HAINES, "DIRTY BOMBS": REASON TO WORRY? 3 (July 2014) http://www.fpri.org/docs/haines_-_dirty_bombs_0.pdf ("Plutonium-238 might be used in an eRDD because of the biological hazards from inhaling alpha particles; however, a malefactor would likely find it easier to obtain other radionuclides."); *see also* Press Release, Nat'l Nuclear Sec. Admin., NNSA Partners with Russia to Recover Material that Could be Used in Dirty Bombs (Nov. 7, 2013), *available at* http://nnsa.energy.gov/mediaroom/pressreleases/rtg (discussing joint efforts of the Department of Energy's National Nuclear Security Administration and the Russian Federation in removing from Arctic maritime locations RTGs containing Strontium-90, "a high-activity radioisotope that could be used in a dirty bomb").

42. Haw. Cnty. Green Party v. Clinton, 980 F. Supp. 1160, 1162 (D. Haw. 1997).

43. According to some references, the campaign referred to itself in all capital letters with an exclamations point: "STOP CASSINI!"

44. *See* Victoria Pidgeon Friedensen, Protest Space: A Study of Technology Choice, Perception of Risk, and Space Exploration 26 (Oct. 11, 1999) (unpublished M.S. thesis, Virginia Polytechnic Institute and State University), *available at* http://scholar.lib.vt.edu/theses/available/etd-120899-134345/.

phase of the flight.[45] NASA argued that the hardening, along with the low probability of a rocket failure, meant that the Cassini mission did not pose a significant risk.[46]

NASA had learned from experience about the desirability of hardening the RTG components. In April 1964, NASA aborted the launch of a Transit 5 navigational satellite, carrying an RTG called SNAP 9A with 2.2 pounds of plutonium-238.[47] Falling back to Earth, the RTG burned up and disintegrated, releasing its plutonium into the upper atmosphere.[48]

Back in 1964, disintegration was actually NASA's intended "failure mode" in the event of a launch abort. The thinking at the time was that it was better to spread out the material globally rather than have it affect a specific population.[49] Years later, the government conducted follow-up research taking over 60 soil samples from around the world.[50] The resulting report found that the SNAP 9A plutonium wound up all over the planet, spread out over both the southern and northern hemispheres.[51] Because the radioactive material was dispersed globally, and because cancer is typically latent for a period of many years, whatever cancers may have resulted from this event are not statistically separable from background cancer rates. That means that whatever deaths resulted—and, as a statistical matter, there were many[52]—they are not traceable to SNAP 9A or NASA.[53]

After the SNAP 9A mishap, redesigned RTGs were hardened to survive rocket failure.[54] NASA had two additional accidents with RTGs, but both landed on the ocean floor intact.[55]

All uses of RTGs in spaceflight entail questions about what would happen in the event of a launch failure. But on its seven-year journey to

45. *Id.* at 27.

46. *Id.* at 28.

47. MARIETTA BENKÖ ET AL., SPACE LAW IN THE UNITED NATIONS 54 (1985).

48. *Id.*

49. *See* Friedensen, *supra* note 44, at 31 (citing NASA, FINAL ENVIRONMENTAL IMPACT STATEMENT FOR THE CASSINI MISSION viii–ix (1995), http://saturn.jpl.nasa.gov/spacecraft/safety/feis.pdf).

50. EDWARD P. HARDY, JR. ET AL., GLOBAL INVENTORY AND DISTRIBUTION OF PU-238 FROM SNAP-9A (1972), http://www.osti.gov/scitech/servlets/purl/4689831.

51. *Id.* at 17.

52. *See infra* note 76 and accompanying text.

53. *See* HARDY, *supra* note 50, at 17.

54. *See* Friedensen, *supra* note 44, at 31; NASA, NASA FACTS: SAFETY OF RADIOISOTOPE POWER SYSTEMS 2 (2012), https://solarsystem.nasa.gov/rps/docs/FINAL_APPRPSSafetyFactSheet052112.pdf.

55. In May 1968, a Nimbus weather satellite launch was aborted shortly after liftoff and the intact RTG was recovered from the seafloor. Then, in April 1970, the Apollo 13 manned mission to the moon was aborted and the lunar module *Aquarius* was used as a lifeboat to keep the mission's three astronauts alive on the way back to Earth. *Aquarius* carried an RTG. As a result, instead of being left on the moon, the RTG re-entered the Earth's atmosphere with *Aquarius.* According to NASA, the RTG remains in deep water in the Pacific and no associated radiation release has been detected. *See Radioisotope Power Systems: Safety*, NASA, https://solarsystem.nasa.gov/rps/safety.cfm (last visited Jan. 24, 2016).

Saturn, Cassini had something else in its mission plan that raised eyebrows. After leaving Earth, Cassini was to fly by various planets on its way to Saturn in order to undertake a number of "gravity assist maneuvers." The maneuver, also known as a "swing-by" or "gravitational slingshot," allows a spacecraft to pick up speed by falling toward a planet and then careening off in a different direction.[56] The spacecraft get a net boost in speed by taking a tiny fraction of the planet's momentum. Sticklers for physics, NASA notes the moniker "slingshot" is not quite accurate. In a description that seems ill-calibrated to soothe nerves, NASA explains, "Gravity assist is really much more like a ping-pong ball hitting the revolving blade of a ceiling fan than it is like a slingshot."[57]

On its way to Saturn, Cassini was scheduled to make four planetary swing-bys. The first two would take it by Venus. The third, scheduled for the summer of 1999, was for Earth. It was the Earth encounter that alarmed critics. Already sped up by Venus, Cassini was to zoom toward Earth at 42,300 miles per hour and pass within 500 miles of the Earth's surface—close enough to be well within the orbits of satellites—before zipping off to the outer solar system.[58] The fear was that if Cassini were even slightly off its trajectory, it could burn up in the Earth's atmosphere, turning its plutonium power supply into a fine oncogenic mist that would travel around the world.[59]

NASA admitted that the hardening of the plutonium supply on Cassini was not sufficient to survive such an atmospheric reentry, but NASA nonetheless found no cause for worry. Specifically, NASA found the chance of such an event to be less than one in a million. NASA's environmental impact statement estimated that in the event Cassini disintegrated in a fly-by mishap, of the 5 billion people exposed to atmospheric plutonium-238, about 1 billion would die of cancer from some other cause, and the deaths attributable to Cassini plutonium would "represent an additional 0.0005 percent above the normally observed cancer fatalities."[60] Translated out of mathematical abstruseness, NASA was predicting 5,000 additional deaths attributable to a Cassini swing-by accident.[61]

56. NASA Jet Propulsion Lab., *Cassini Solstice Mission: Gravity Assists/Flybys A Quick Gravity Assist Primer*, NASA, http://saturn.jpl.nasa.gov/mission/missiongravityassistprimer/ (last visited Jan. 24, 2016).

57. *Id.*

58. *See* Karl Grossman, *The Risk of Cassini Probe Plutonium*, CHRISTIAN SCI. MONITOR (Oct. 10, 1997), http://www.csmonitor.com/1997/1010/101097.opin.opin.1.html.

59. *See generally id.*

60. *See* Friedensen, *supra* note 44, at 30. It should be noted that Friedensen's thesis cites the figure as 500,000, but a 0.0005% increase over 1 billion deaths from cancer from other causes translates to 5,000 additional deaths. I am grateful to Ms. Friedensen for responding to my inquiry on this. *See also* Grossman, *supra* note 58 (reporting NASA as stating an estimate of 2,300 fatalities over a 50-year period).

61. *Id.*

While 5,000 deaths may seem like a lot, many observers thought the estimate was undeservedly sanguine. One critic was Michio Kaku, a professor of physics at City University of New York and author of several popular science books. He said, "I find that NASA bureaucrats in some sense are living in Fantasyland . . . . Pure guesswork has replaced rigorous physics. Many of these numbers are simply made up."[62] Kaku thought 200,000 deaths was a fairer estimate.[63] Other estimates were even higher. John Gofman, an emeritus professor of molecular and cellular biology at the University of California, Berkeley, estimated up to 1 million deaths would result from a Cassini swing-by burn-up.[64] And Ernest J. Sternglass, an emeritus professor of radiological physics at the University of Pittsburgh School of Medicine, suggested 40 million deaths could result.[65]

Cassini launched on October 15, 1997.[66] In August 1999, Cassini completed its gravitational slingshot around Earth without incident.[67]

Although the controversy vanished from the public eye, there is a fascinating, unwritten epilogue. To come up with its less-than-one-in-a-million probability for a Cassini swing-by accident, NASA considered accident scenarios such as a micrometeoroid hitting a propellant tank.[68] But NASA assigned virtually no weight to the possibility of Cassini going off track because of software errors or navigational design errors. NASA estimated the probability of those occurrences at 3 in 1 trillion and 7 in 10 billion, respectively.[69] Yet in December 1998, well after Cassini was launched but before its Earth swing-by, NASA launched its Mars Climate Orbiter, a spacecraft intended to function as a kind of weather satellite for the red planet. Just weeks after Cassini flew by Earth without incident, NASA lost contact with the Mars Climate Orbiter.[70] A subsequent investigation determined that the spacecraft burned up in the Martian atmosphere. The reason the Mars Climate Orbiter came in too low was that some of its software code used English unit measurements instead of the metric system[71]—the kind of failure that NASA thought so improbable for Cassini as to be statistically insignificant.

---

62. *Cassini Roars into Space,* CNN (Oct. 15, 1997), http://edition.cnn.com/TECH/9710/15/cassini.launch/.

63. Grossman, *supra* note 58.

64. *Id.*

65. *Id.*

66. NASA Jet Propulsion Lab., *Cassini Solstice Mission: Quick Facts*, NASA, http://saturn.jpl.nasa.gov/mission/quickfacts/ (last visited Jan. 24, 2016).

67. *See id.*

68. NASA, FINAL ENVIRONMENTAL IMPACT STATEMENT FOR THE CASSINI MISSION app. at B-9 (1995), *available at* http://saturn.jpl.nasa.gov/spacecraft/safety/appendb.pdf.

69. *Id.*

70. MARS CLIMATE ORBITER MISHAP INVESTIGATION BD., PHASE I REPORT 10 (1999), *available at* ftp://ftp.hq.nasa.gov/pub/pao/reports/1999/MCO_report.pdf.

71. *Id.* at 10, 16.

One imagines that if the Mars Climate Orbiter fiasco had happened before the Cassini Earth swing-by, NASA might have felt some public pressure to re-evaluate its RTG program. The sequence of events being what it was, public concern over RTGs did not resurface. Today, NASA's use of RTGs has expanded. NASA's most recent robotic rover on Mars, for instance, uses an RTG, departing from the past practice of outfitting Mars rovers with solar panels.[72]

NASA remains upbeat about RTG safety. A 2012 NASA fact sheet on RTGs, in a section titled, "A Long Record of Success," explains that "NASA has an outstanding record of safety in launching [RTGs], with 17 successful launches and no failures over the past three decades."[73] The document does acknowledge that there were incidents before 1971. But, it points out, "[i]n every instance, the radioisotope power system performed as it was designed"[74]—a characterization that puts a rosy gloss on the SNAP 9A plutonium dispersal in 1964. The fact sheet also states that "[n]o member of the public has ever been injured in a NASA launch."[75] Yet if one extrapolates from NASA's projection of 5,000 deaths from a Cassini burn-up, 152 deaths can be attributed to the SNAP 9A dispersal.[76]

My aim in reviewing the Cassini plutonium issue is not to make an argument against NASA's use of RTGs. Nor would I opine that NASA's RTGs are acceptably safe. The question I am interested in is: *Who should decide?* By reviewing the Cassini plutonium case, I hope to have provided an example that will illustrate my discussion, further below, of the potential pitfalls of having agencies decide for themselves whether their activities constitute an acceptable risk to the public. The disaster scenario I discuss next is another such example.

---

72. NASA's solar-powered Opportunity and Spirit rovers landed on Mars in 2004 for a three-month mission; Spirit continued working for six years and Opportunity is still active as of 2015. *See* NASA, *Opportunity Mars Rover Preparing for Active Winter*, NASA (Sept. 25, 2015), http://mars.nasa.gov/news/whatsnew/index.cfm?FuseAction=ShowNews&NewsID=1857 (discussing latest status); *Mars Exploration Rovers: Technology*, NASA (detailing rovers' solar power system). NASA's plutonium-powered Curiosity rover landed in 2012 for a 23-month primary mission and is still active as of 2015. *See* NASA, NASA FACTS: MARS SCIENCE LABORATORY/CURIOSITY 1, http://mars.nasa.gov/msl/news/pdfs/MSL_Fact_Sheet.pdf (discussing mission generally); NASA, NASA FACTS: MARS EXPLORATION: RADIOISOTOPE POWER AND HEATING FOR MARS SURFACE EXPLORATION 1–2, http://www.jpl.nasa.gov/news/fact_sheets/mars-power-heating.pdf (describing advantages of radioisotope power); NASA, *NASA Mars Rover Curiosity Reaches Sand Dunes*, NASA (Dec. 11, 2015), http://mars.nasa.gov/msl/news/whatsnew/index.cfm?FuseAction=ShowNews&NewsID=1876 (discussing latest status).

73. NASA, SAFETY OF RADIOISOTOPE POWER SYSTEMS, *supra* note 54, at 2.

74. *Id.*

75. *Id.*

76. This extrapolation is from the relative masses of plutonium-238 on Cassini and SNAP 9A. *See supra* notes 42, 47 and accompanying text.

### C. *Brookhaven Laboratory and Strange-Matter Conversion*

On Long Island, about an hour's drive east from New York City, the DOE's Brookhaven National Laboratory operates a particle accelerator called the Relativistic Heavy Ion Collider ("RHIC," pronounced "Rick"). The aim of the RHIC is to replicate the state of the universe in the ultrahot instant after the Big Bang.[77]

Some expressed concern, however, about the RHIC's venture into unknown realms of physics—particularly a question of whether the experiment might create a "strangelet," a tiny particle of exotic strange matter.[78] Creating a strangelet would be a triumph of modern physics. In an unlikely scenario, however, it might also be unbelievably dangerous—unstoppably transforming and absorbing all normal matter it touches. After a latency of many years, the concern is, the accreting mass of strange matter within the Earth would overtake the whole planet. In the words of one eminent scientist, the Earth would be left "an inert hyperdense sphere about one hundred metres across."[79]

The RHIC works by taking atoms of heavy elements—routinely gold—stripping off the electrons, and then introducing the bare nuclei—or ions—into a ring of supercooled magnets 2.4 miles around.[80] Ion beams circulate in two different directions. One ion beam goes clockwise, the other goes counterclockwise.[81] The ions are propelled around and around with increasing amounts of energy until each is traveling 99.995% of the speed of light.[82] Then, at crisscross points along the accelerator's circumference, the nuclei come together in head-on collisions.[83] The colliding ions produce incredibly hot temperatures—reaching 4 trillion degrees Celsius.[84] By comparison, the superhot core of the sun is a quarter-million times cooler.[85]

The heat is enough to tear apart the nuclei's constituent neutrons and protons, creating a plasma of elementary particles known as quarks and gluons.[86] It is this "primordial plasma" that replicates the immediate aftermath of the Big Bang.[87] The original aim of the RHIC program was to see if this plasma could be produced. That goal having been achieved, the program's aim now is gathering detailed data on the plasma's properties.[88]

---

77. *Big Chill Sets in as RHIC Physics Heats Up*, BROOKHAVEN NAT'L LABORATORY (Feb. 3, 2014), http://www.bnl.gov/newsroom/news.php?a=11606 [hereinafter *RHIC Big Chill*].

78. I explain this in detail below. *See infra* note 95 and accompanying text.

79. MARTIN REES, OUR FINAL HOUR: A SCIENTIST'S WARNING: HOW TERROR, ERROR, AND ENVIRONMENTAL DISASTER THREATEN HUMANKIND'S FUTURE IN THIS CENTURY—ON EARTH AND BEYOND 121 (2003).

80. *The Physics of RHIC*, BROOKHAVEN NAT'L LABORATORY, http://www.bnl.gov/rhic/physics.asp (last visited Jan. 24, 2016) [hereinafter *RHIC Physics*].

81. *Id.*

82. *RHIC by the Numbers*, BROOKHAVEN NAT'L LABORATORY, http://www.bnl.gov/bnlweb/pubaf/fact_sheet/pdf/fs_rhic_numbers.pdf (last visited Jan. 24, 2016).

83. *RHIC Physics*, *supra* note 80.

84. *RHIC Big Chill*, *supra* note 77.

85. *Id.*

86. *Id.*

The disaster scenario has to do with what happens after these primordial particles are liberated. Once freed from their ordinary, current-universe form, the worry is that the quarks and gluons might, upon cooling, coalesce into the theorized configuration of strange matter.[89]

The strangelet question might never have been given a public airing. Brookhaven National Laboratory certainly did not advertise the issue. But in an ironic twist, the public concern about strange matter seems to have been triggered by a physicist trying to tamp down fears of a black hole calamity.

In 1999, *Scientific American* magazine published an article about the RHIC's forthcoming launch.[90] That article touched off a flurry of letters from alarmed readers who thought RHIC experiments might lead to the creation of black holes.[91] That, in turn, prompted the magazine to get physicist Frank Wilczek to respond.[92] Wilczek, who was later awarded the Nobel Prize,[93] wrote that the RHIC was not capable of producing black holes, so there was no need to worry on that score.[94]

Then, seemingly apropos of nothing, Wilczek added:

> On the other hand, there is a speculative but quite respectable possibility that subatomic chunks of a new stable form of matter called strangelets might be produced (this would be an extraordinary discovery). One might be concerned about an "ice-9"-type transition, wherein a strangelet grows by incorporating and transforming the ordinary matter in its surroundings. But strangelets, if they exist at all, are not aggressive, and they will start out very, very small. So here again a doomsday scenario is not plausible.[95]

"Ice-9" refers to the science-fiction brainchild of Kurt Vonnegut: a form of ice that is solid at room temperature and is capable of converting any liquid water it touches into more solid ice-9. In Vonnegut's novel

87. *Id.*

88. *Id.*

89. Sheldon L. Glashow & Richard Wilson, *Taking Serious Risks Seriously*, 402 NATURE 596, 596–97 (1999).

90. Madhusree Mukerjee, *A Little Big Bang*, SCI. AM, Mar. 1999, at 60.

91. Walter L. Wagner, Letter to the Editor, *Black Holes at Brookhaven?*, SCI. AM., July 1999, at 8.

92. Frank Wilczek, Letter to the Editor, *Black Holes at Brookhaven?*, SCI. AM., July 1999, at 8.

93. *The Nobel Prize in Physics 2004*, NOBELPRIZE.ORG, http://www.nobelprize.org/nobel_prizes/physics/laureates/2004/ (last visited Jan. 24, 2016).

94. Wilczek, *supra* note 92, at 8. Wilczek's theoretical assumptions for why black holes could not be produced by present-era particle accelerators turned out later to be undermined by advances in string theory. *See* Eric E. Johnson, *The Black Hole Case: The Injunction Against the End of the World*, 76 TENN. L. REV. 819, 838–40 (2009).

95. Wilczek, *supra* note 92, at 8.

*Cat's Cradle*, ice-9 sets off a doomsday chain reaction, freezing the world's oceans and rivers and exterminating plant and animal life by freezing the water inside cells.[96] Not exactly a reassuring analogy.

Despite Wilczek's proffered opinion that a strangelet disaster was "not plausible,"[97] his comments generated a surge of media inquiries. In a subsequent treatment in the journal *Science*, Harvard physicists Sheldon L. Glashow and Richard Wilson framed the strangelet question this way:

> If strangelets exist (which is conceivable), and if they form reasonably stable lumps (which is unlikely), and if they are negatively charged (although the theory strongly favours positive charges), and if tiny strangelets can be created at RHIC (which is exceedingly unlikely), then there just might be a problem. A newborn strangelet could engulf atomic nuclei, growing relentlessly and ultimately consuming the Earth.[98]

As Glashow and Wilson's parentheticals attest, the strangelet risk question is bedeviled by uncertainty. Strange matter has yet to be observed. It remains, at present, a theoretical construct. (Indeed, quark-gluon plasma itself was purely theoretical until it was successfully produced and observed at the RHIC.)[99] Because strange matter has not been observed, its particular properties are not known. And the particular properties of strangelets are crucial, because some theoretically possible versions of strangelets would be entirely harmless—including, for instance, strangelets with quick decay times[100] or strangelets with positive electrical charge.[101] But a strangelet with certain properties of stability and negative charge could initiate a chain reaction.[102]

One team of physicists offered this analysis of the end of such a process:

> [G]ravity and thermal motion may then sustain the accreting chain reaction until, perhaps, the whole planet is digested, leaving behind a strangelet with roughly the mass of the Earth and ~ 100 m radius. The release of energy per nucleon should be of the order of

---

96. *See* KURT VONNEGUT, CAT'S CRADLE 46 (2010 Dial Press ed.) (1963).

97. Wilczek, *supra* note 92, at 8.

98. Glashow & Wilson, *supra* note 89, at 597.

99. John Matson, *Nuclear Decelerator: Last U.S. Particle Collider on Chopping Block*, SCI. AM. (Aug. 24, 2012), http://www.scientificamerican.com/article/rhic-jlab-frib-budget-cuts/.

100. If strangelets decay quickly enough, they will cease to exist before they can create a self-sustaining chain reaction. *See* Glashow & Wilson, *supra* note 89, at 597.

101. A strangelet with positive charge would be safe, it is argued, because it would be repulsed by the positively charged nuclei in everyday matter, and thus would not convert those nuclei. *See* Kate Wong, *The Safety of Strangelets*, SCI. AM. (Nov. 22, 2000), http://www.scientificamerican.com/article/the-safety-of-strangelets/.

102. Glashow & Wilson, *supra* note 89, at 597.

several MeV and, if the process is a run-away one, the planet would end in a supernova-like catastrophe.[103]

Brookhaven's director convened a committee of four physicists—Wit Busza, Robert L. Jaffe, and Frank Wilczek of MIT, plus Jack Sandweiss of Yale—to produce a report on theorized disaster scenarios for the RHIC.[104]

One of the scientists on the committee called the safety question "absurd" and candidly suggested that the necessity for the report sprang from the need to guard the laboratory's reputation.[105] In other words, the main problem from Brookhaven's perspective seemed to be one of public relations.[106] Yet not all scientists saw it that way. Far from characterizing the issue as absurd, Glashow and Wilson wrote, "It is a fair concern: one that must be raised[.]"[107]

The Busza panel finished its work by the end of September 1999 and issued its findings. The report concluded that a strangelet catastrophe was "firmly excluded" on the basis of theory.[108] On top of that, the panel said that empirical evidence allowed them to "decisively rule out" the possibility of a dangerous strangelet.[109] Any delay in starting up the RHIC, according to the report, was not warranted.[110]

Around the same time, a group of European physicists at the international CERN laboratory also looked into the killer strangelet scenario. Their interest in the matter may have been prompted by the fact that CERN was planning to operate a similar experiment a few years hence. So, a delay in the experimental program at RHIC could well have redounded to CERN. The team of three authors—Arnon Dar, Alvaro de Rújula, and Ulrich Heinz—reached the same bottomline assessment as the Busza panel. They concluded that the RHIC posed no danger.[111] Their paper, published in October 1999, included the assurance that "our extremely conservative conclusion is that it is safe to run RHIC for 500 million years."[112]

---

103. Arnon Dar, A. De Rújula, & Ulrich Heinz, *Will Relativistic Heavy-ion Colliders Destroy Our Planet?*, 470 PHYSICS LETTERS B 142, 143 (1999). As noted below, this team of authors concluded the RHIC is safe. *See infra* notes 111–112 and accompanying text.

104. BUSZA ET AL., *supra* note 5, at 1.

105. Robert Jaffe, quoted in Erik Snowberg, *Robert Jaffe: Academics, Activism, and Armageddon*, TECH, Aug. 4, 1999, http://tech.mit.edu/V119/PDF/V119-N30.pdf (Jaffe: "As a consequence of the safety issues that were ranged, the director of Brookhaven is in a very difficult position. . . . If someone raises an absurd safety question, he has two options. He can ignore it, or he can take it seriously and go at it like a typical science problem. His feeling was that if he ignored this, the accusation that he was not taking safety seriously would have been too damaging to the reputation of the laboratory. He feels he must take it seriously. However, by taking it seriously, he is also legitimizing it. People say, 'This must be a problem because, after all, the director of Brookhaven appointed a panel to look at it.'").

106. Blogger Eliezer Yudkowsky put it this way: "[T]he RHIC Review is a work of public relations. . . . Everyone knew, before the RHIC report was written, what answer it was supposed to produce. That is a very grave matter. Analysis is what you get when physicists sit down together and say, 'Let us be curious,' and walk through all the arguments they can think of, recording them as they go, and finally weigh them up and reach a conclusion." Eliezer Yudkowsky, *LA-602 vs. RHIC Review*, LESS WRONG (June 19, 2008, 10:00 AM), http://lesswrong.com/lw/rg/la602_vs_rhic_review/. Yudkowsky added that he is not personally concerned about particle-accelerator safety. *Id.*

107. Glashow & Wilson, *supra* note 89, at 596.

108. BUSZA ET AL., *supra* note 5, at 19.

109. *Id.* at 10.

110. *Id.* at 1.

A fresh round of media attention covered the reassuring results.[113] In the December 1999 issue of *Science,* Glashow and Wilson wrote that the Busza and Dar papers "answered decisively" the safety question.[114] Yet like the Cassini controversy, the RHIC story has a little-noticed epilogue.

After the media stories ebbed away, once third parties had time to digest the content of the reports, a number of critics came forth with reasons to doubt the steadfastness of the conclusion that the RHIC is safe.

One of the first critics, who emerged the following year, was Francesco Calogero of the University of Roma–La Sapienza.[115] As head of the Pugwash Conferences on nuclear disarmament, Calogero accepted the Nobel Peace Prize on behalf of his organization.[116] While Calogero thought the overall probability of a strangelet disaster was "tiny,"[117] he nonetheless identified a number of troubling issues with the published reports. In terms of specifics, Calogero set out detailed objections to safety analyses in both the Dar and Busza papers.[118] On a more general level, Calogero criticized the RHIC's safety assurances as lacking in candor.[119] He wrote that the physicists involved seemed "more concerned with the public relations impact . . . than in making sure that the facts are presented with complete scientific objectivity."[120]

Calogero also pointed out a couple of aspects of the Dar and Jaffe analyses overlooked by the media when writing stories about the safety assurances: The Dar paper laid bare an important problem in the Busza model used to exclude a strangelet disaster on empirical grounds.[121] And a revised version of the Busza analysis pointed out that under certain assumptions, the probability ceiling for disaster calculated by the Dar pa-

111. *Id.* at 148.
112. *Id.* at 146.
113. *See, e.g.*, Curt Suplee, *Scare Stories and Mysteries of Quarky Behavior*, WASH. POST (Sept. 13, 1999), http://www.washingtonpost.com/archive/politics/1999/09/13/scare-stories-and-mysteries-of-quarky-behavior/4fb880b4-48a0-4df7-996f-11d64291746e/.
114. Glashow & Wilson, *supra* note 89, at 596.
115. Francesco Calogero, *Might a Laboratory Experiment Destroy Planet Earth?*, 25 INTERDISC. SCI. REVS. 191, 202 (2000).
116. *Id.*
117. *Id.* at 191.
118. *See id.* at 195–96.
119. *Id.* at 198.
120. *Id.*
121. *See id.* at 195–96.

per evaporates.[122] Thus, although the Dar paper and Busza paper both agreed that the RHIC was safe, each paper perceived a substantial flaw in the other's analysis.

Another critic was H. Kimball Hansen, an emeritus professor of astronomy with Brigham Young University. Hansen submitted an affidavit in a lawsuit against the RHIC.[123] He criticized the Dar paper's reliance on observations of astronomical supernovae as a means of providing an empirical foundation for the safety assessment, calling the analysis "wholly faulty."[124]

The highest profile critic was Sir Martin Rees, Astronomer Royal of the United Kingdom, who wrote a book in which he said the theorists "seemed to have aimed to reassure the public . . . rather than to make an objective analysis."[125]

The most detailed critique came from Adrian Kent, a Cambridge University theoretical physicist.[126] Kent wrote a paper focused on the Dar and Busza claims that catastrophic risk at the RHIC could be excluded on empirical grounds.

To understand Kent's critique, it is helpful to pause for a moment and consider how it is that the RHIC's safety could be empirically demonstrated at all. Empirical evidence requires data from a sample set, and that is generally done through repeated trials. For instance, if you want to empirically demonstrate that a drug is safe, you administer the drug to a limited group of people and then statistically analyze the results.[127] But you can't administer the RHIC to a limited number of people, since the question is whether the machine threatens human extinction. So how do you get empirical data on whether RHIC can be safely run before you've run the RHIC?

The technique used by both the Busza and Dar papers was to use a naturally occurring process as a stand-in for the RHIC's ion collisions. Both papers looked to cosmic rays—the name given to independent particles that speed through the universe at very close to the speed of light. Some cosmic ray particles are heavy ions, which are similar to the sped-

122. *See id.* at 196; *see also* R.L. Jaffe, W. Busza, J. Sandweiss, & F. Wilczek, *Review of Speculative "Disaster Scenarios" at RHIC*, ARXIV:HEP-PH/9910333v2, May 19, 2000, at 6 [hereinafter Jaffe Version 2].

123. Affidavit of H. Kimball Hansen, Wagner v. U.S. Dep't of Energy, No. C99-2226 (N.D. Cal. filed May 18, 2000).

124 *Id.* at 2.

125. REES, *supra* note 79, at 127.

126. *See* Kent, *Critical Look*, *supra* note 3. Kent's *Critical Look* paper was first posted to the arXiv, an online research paper repository, in September 2000. For the posting and revision history, see http://arxiv.org/abs/hep-ph/0009204.

127. *See, e.g.*, Stuart R. Cohn & Erin M. Swick, *The Sitting Ducks of Securities Class Action Litigation: Bio-Pharmas and the Need for Improved Evaluation of Scientific Data*, 35 DEL. J. CORP. L. 911, 916–19 (2010) (explaining the FDA drug approval process).

up ions produced by the RHIC.[128] The Dar paper constructed a model that involved cosmic rays hitting each other.[129] The Busza paper counted cosmic rays hitting the moon.[130] The Busza group's argument, for instance, is that if RHIC collisions could generate a dangerous strangelet that could eat the Earth, then naturally occurring cosmic-ray collisions with the lunar soil would likely have destroyed the moon by now via the same process.[131]

Among the models the Busza paper considered was one taking account of cosmic rays composed of iron nuclei hitting the moon at RHIC-level energies. Based on how many such cosmic rays hit the moon and how long the moon has existed, the Busza paper found it is unlikely the moon would still be intact if it were possible for strangelets to be formed under RHIC-like conditions—so unlikely that the model "leaves a comfortable margin of error."[132]

One of the helpful aspects of Kent's work is the translation of Busza's quantitative results into a form more easily understood by lay readers. Kent explained that the "probability bound"—meaning the maximum-possible risk[133]—implied by Busza's analysis was approximately $10^{-4}$; translated out of scientific notation, that means the Busza authors found there was no more than a one-in-10,000 chance that the RHIC would destroy the Earth.[134] It was this result the Busza report deemed "comfortable."

---

128. BUSZA ET AL., *supra* note 5, at 8 ("Cosmic ray processes accurately reproduce the conditions planned for RHIC. They are known to include heavy nuclei and to reach extremely high energies.").

129. *See* Dar et al., *supra* note 103.

130. Why did the Busza paper use the moon, instead of the Earth? The answer is that the moon is without an atmosphere that protects it from cosmic-ray collisions. *See, e.g.*, BUSZA ET AL., *supra* note 5, at 9 (referring to the moon's lack of an atmosphere); Dar et al., *supra* note 103, at 144 (referring to the protection of a light-gas atmosphere).

131. *See* BUSZA ET AL., *supra* note 5.

132. *Id.* at 23.

133. It's important to distinguish a probability bound from a probability. A probability bound of 1-in-10,000 does not mean that something has a one-in-10,000 chance of happening. It means, rather, that the chance of it happening *is not greater than* one-in-10,000 chance of happening. Assuming the probability bound is correct, it implies that the real risk could be zero, or it could be any number up to one-in-10,000.

134. In a previous paper, I quoted an erroneous figure (one in 5,000) for this probability. *See* Johnson *supra* note 94, at 896, 906. There, I had taken the number from Kent's paper, in its published form. *See* Adrian Kent, *A Critical Look at Risk Assessment for Global Catastrophes*, 24 RISK ANALYSIS 157, 161 (2004) (quoting "a risk bound of one in 5,000"). In doing follow up research a few years later, I tried independently to derive the same figure from the Busza paper, and I could not. I contacted Dr. Kent, who was kind enough to look into the matter. He subsequently posted an updated version of the paper in July 2015 correcting the figure to one-in-10,000. *See* Kent, *Critical Look*, *supra* note 3, at 12 (discussing correction and updating). In making the correction, Kent provides additional analysis about the Busza group's probability bounds. Referencing the second version of the Busza paper, Kent writes: "BJSW's figures are consistent with a joint probability of both the Moon surviving to date and a hypothetical catastrophe at RHIC of $e^{-1}$ times the stated catastrophe risk bounds. Since the probability of hypothetical catastrophe at RHIC is at least as large as this joint probability, it follows that (however one treats the separate events of Moon survival and hypothetical RHIC catastrophe) no catastrophe risk bound better than $e^{-1}$ times the stated figures can be derived from BJSW's calcula-This idea of acceptable risk would seem to be unique, to say the least. A one-in-10,000 risk of causing *one individual death* is commonly deemed unacceptable by regulatory agencies such as the FDA and EPA, and action by these agencies may be triggered by just a one-in-a-million chance of causing a single death.[135]

Kent suspected "some surprising confusion on [Busza and co-authors'] part at the time of writing," since, in his view, "no sane person would seek to reassure the public by suggesting that a risk bound of 1 in ≈ 10000 of destroying the Earth represented a comfortable margin of error."[136]

In fact, once Kent communicated this criticism to the Busza team, those authors backed off of their original statement. In a revised version of the paper, the Busza authors refined their calculations to add an extra order of magnitude, putting a ceiling on the risk of disaster for the relevant model at one in 100,000.[137] They also removed the word "comfortable" and said they would not attempt to define acceptable risk.[138]

Kent also identified a quite astounding conceptual mathematical error made by the Dar authors.

---

tions without introducing further assumptions that BJSW did not suggest in this part of their discussion. In particular, on BJSW's most conservative assumptions, no risk bound better than 1 in 36788 can be derived." *Id.* (citing Jaffe Version 2, *supra* note 122). Applying this reasoning to the first version of the Busza paper, the first version supported a risk bound of no better than approximately one in 3,679. Compare note 137, *infra*, and accompanying text discussing the probability bound change from one-in-10,000 to one-in-100,000.

135. *See, e.g.*, Matthew D. Adler, *Against "Individual Risk": A Sympathetic Critique of Risk Assessment*, 153 U. PA. L. REV. 1121, 1122–23 (2005) (use of one-in-a-million threshold in EPA regulatory contexts of air pollution, water pollution, and pesticide regulation; EPA acceptable risk range for Superfund cleanup between one in 10,000 and one in 1 million for lifetime fatality risk for individuals with maximal exposure; FDA's traditional threshold of one in 1 million for carcinogenic constituents of food); Adam Babich, *Too Much Science in Environmental Law*, 28 COLUM. J. ENVTL. L. 119, 152–53 (2003) (EPA's acceptable risk range can be from a one-in-a-million to one-in-10,000 chance of an individual death, with one-in-a-million being EPA's historical threshold and remaining the starting point for analysis); Listing of D&C Orange No. 17 for Use in Externally Applied Drugs and Cosmetics, 51 Fed. Reg. 28,331, 28,345 (Aug. 7, 1986) (FDA found "1 in 1 million level has become a benchmark in the evaluation of the safety of carcinogenic compounds administered to food-producing animals"). Note, however, that in some contexts, agencies may find considerably higher levels of risk to be acceptable, such as where the benefits are seen to outweigh the risk. *See, e.g.*, Frank B. Cross, *Beyond* Benzene*: Establishing Principles for A Significance Threshold on Regulatable Risks of Cancer*, 35 EMORY L.J. 1, 43 (1986) (given the benefits of nuclear power, the Nuclear Regulatory Commission considered an increased chance of cancer from plant accidents of 1.3 in 10,000).

136. *See* Kent, *Critical Look*, *supra* note 3, at 5.

137. *See* R.L. Jaffe, W. Busza, J. Sandweiss, & F. Wilczek, *Review of Speculative "Disaster Scenarios" at RHIC*, 72 REV. MOD. PHYSICS 1125, 1138 (2000) (specifying a bound of $10^{-5}$ for the model called "case II"); Jaffe Version 2, *supra* note 122, at 24 (same).

138. See Jaffe Version 2, *supra* note 122, at 3 ("We do not attempt to decide what is an acceptible upper limit on p, nor do we attempt a 'risk analysis', weighing the probability of an adverse event against the severity of its consequences."); Kent, *Critical Look*, *supra* note 3, at 3 ("[T]he claim was withdrawn by BJSW, after criticisms from the author of this paper. BJSW produced a second version of their preprint, removing the reassuring characterisations of their risk bounds and instead disavowing any attempt to decide what is an acceptable upper bound on [the probability of catastrophe].").

As mentioned above, the Dar team stated: "[O]ur extremely conservative conclusion is that it is safe to run RHIC for 500 million years."[139] Yet that is not what the Dar team's analysis showed. Their analysis indicated a probability bound of a one-in-500-million chance that the RHIC could create a planet-eating strangelet in any given year.[140] As Kent pointed out, that does not mean that it would be safe to run the RHIC for 500 million years, since such a probability bound is consistent with a high probability of destroying the Earth within that time.[141] The Dar authors' statement is like saying it would be safe to keep your house in a floodplain for 100 years, even if the floodplain is subject to a once-in-a-century risk of flooding.

Then, according to Kent, the Busza team made their own conceptual error in describing the Dar paper's analysis. The Busza paper stated that the Dar result was "a factor of $10^8$ below the value required for the safety of RHIC."[142] But that cannot be right. Kent explained, "[A] risk bound $10^8$ times that of [the Dar paper's] would be consistent with a high probability of destroying the Earth within 5 years of the RHIC experiment—a risk level which even the most gung-ho physicist could hardly describe as 'safety.'"[143]

It's worth emphasizing that these problems with the Busza and Dar safety analyses do not imply that the RHIC poses a high-probability risk of destroying the Earth. But such problems do seem to call into question the reliability of these safety analyses. Along these lines, Kent is right in saying that Busza and Dar's "mischaracterisations . . . illustrate very clearly[] that scientists whose expertise is not in risk analysis or public policy cannot necessarily be relied on either to interpret the risk implications of the science correctly or to consider elementary arguments that tend to suggest more cautious risk criteria."[144]

The fact is that the Busza and Dar empirical analyses are built on layers of assumptions—and the reasonability of those assumptions is a judgment call. Given the basic conceptual/mathematical errors in the Dar and Busza papers laid bare by Kent, it seems fair to question how much credence their judgment should be given in terms of the reasonability of the assumptions in their models.[145]

In a separate paper, *Problems with Empirical Bounds for Strangelet Production at RHIC*, Kent listed a series of what he labeled "potential

139. Dar et al., *supra* note 103, at 146.
140. *See* Kent, *Critical Look*, *supra* note 3, at 5.
141. *Id.*
142. BUSZA ET AL., *supra* note 5, at 24.
143. *See* Kent, *Critical Look*, *supra* note 3, at 5.
144. *Id.* at 4.
145. This is my view, not necessarily Kent's.

flaws" with the Dar and Busza analyses.[146] Each are aspects where the Busza or Dar teams used their expert judgment to make assumptions for building their empirical analysis.

As one example, a potential weakness Kent highlights in the Busza analysis was the assumption that collisions of iron nuclei—relatively common in cosmic rays hitting the moon—are sufficiently similar to collisions of gold nuclei—undertaken in the RHIC—such that they are relevant to the issue of RHIC safety.[147] The Busza paper stated that "iron is nearly as good a 'heavy' ion as gold" for safety analysis purposes.[148] The necessity of the assumption is plain: Just as gold is rarer and more precious than iron on Earth, gold is rarer and more precious in the heavens. Cosmic-ray collisions on the moon involving gold vs. gold nuclei happen too infrequently, so there is not enough useful data for any probability bound. That is, if gold-gold collisions have a dangerous-strangelet-producing capacity that iron-iron collisions do not, then the survival of the moon is entirely consistent with a 100% possibility of disaster. As the Busza team acknowledges, "If . . . one insists on recreating exactly the circumstances at RHIC and insists on the worst case rapidity distribution, then lunar limits are not applicable."[149]

For filling in this gap in their safety analysis, the Busza panel refers to the Dar group's analysis. But Kent points out that though the Dar authors "claim their limit is fool-proof, it actually relies on some important assumptions." Kent reviews a number of such assumptions and points out why they might be invalid. One example, that is straightforward to understand, is that the Dar analysis assumes that killer strangelets formed from cosmic-ray collisions in interstellar space would be long-lasting enough to travel along until they met a star and destroyed it through a supernova visible to our telescopes. Kent suggests that if strangelet lifetimes tended to be greater than a millionth of a second but less than a million years, then "strangelets produced at RHIC would still cause catastrophes, while almost no strangelets produced in space would survive long enough to produce observable effects."[150]

Another critic was Judge Richard A. Posner, who looked at the issue in terms of incentives and economics. Posner noted that each of the scientists on the Jaffe panel either had plans to participate in the RHIC program or else wanted the results from the RHIC program to further their own theoretical work.[151] Posner argued that "career concerns can influence judgment in areas of scientific uncertainty, and scientists, like other people, can be overconfident."[152]

---

146. *See* Adrian Kent, *Problems with Empirical Bounds for Strangelet Production at RHIC*, ARXIV:HEP-PH/0009130v2, Sept. 11, 2000, at 6, http://arxiv.org/pdf/hep-ph/0009130v2.pdf [hereinafter Kent, *Problems with Empirical Bounds*].

147. *See id.* at 2–3, 6.

148. BUSZA ET AL., *supra* note 5, at 24. Note that gold nuclei are more than three times as heavy as iron nuclei. The most abundant isotope of gold has 79 protons and 118 neutrons (total of 197 nucleons), and the most abundant isotope of iron has 26 protons and 30 neutrons (total of 56 nucleons). Nat'l Nuclear Data Ctr., Brookhaven Nat'l Lab., Chart of Nuclides, http://www.nndc.bnl.gov/chart/ (last visited Jan. 24, 2016).

149. BUSZA ET AL., *supra* note 5, at 23.

150. Kent, *Problems with Empirical Bounds*, *supra* note 146, at 4.

Posner also offered a cost-benefit analysis suggesting the RHIC was not worth even a small risk of destroying the Earth.[153] He wrote:

> [T]he purpose is to quench scientific curiosity. Obviously, this research benefits scientists, or at least high-energy physicists. But how does such research benefit society as a whole? . . . If there are no good answers, the fact that such research poses even a slight risk of global catastrophe becomes a compelling argument against its continued subsidization.[154]

These criticisms have not had a perceivable impact at Brookhaven, which has continued to run its experiments. In fact, the RHIC program has expanded and evolved since the strangelet controversy was aired.[155]

Originally, the RHIC was scheduled to collide gold ions over a 10-year-long program.[156] Program extensions, however, have kept the RHIC going, and it is now in its 15th year.[157] The program has also changed in ways unanticipated by the Busza team's report. Brookhaven has moved beyond gold nuclei to begin experimenting with copper and uranium ions.[158] The RHIC has also been upgraded to achieve many times more collisions than it was able to make under its original design.[159]

Recently, in responding to questions about upgrades and expansions of the RHIC program, a couple of physicists at Brookhaven suggested that heavy ion collisions completed to this point have had "no issue" thus, "the RHIC program is safe from any of these putative risks."[160] While this seems reassuring, it sweeps under the rug the latency aspect of the strangelet-disaster scenario: As the RHIC's defenders acknowledge, if dangerous strangelets are possible, then it is also possible that, once created, they could remain latent inside the Earth for many years.[161]

---

151. RICHARD A. POSNER, CATASTROPHE: RISK AND RESPONSE 189 (2004).

152. *Id.*

153. *Id.* at 142.

154. Richard A. Posner, *Efficient Responses to Catastrophic Risk*, 6 CHI. J. INT'L L. 511, 521 (2006).

155. *See, e.g.*, Satoshi Ozaki & Thomas Roser, *Relativistic Heavy Ion Collider, Its Construction and Upgrade*, PROGRESS THEORETICAL & EXPERIMENTAL PHYSICS (Jan. 12, 2015), http://ptep.oxfordjournals.org/content/2015/3/03A102.full?sid=ae067b9d-5563-499b-ad5d-e697f26279df ("[T]he scientific objective of RHIC was expanded to include the study of the spin structure of nucleons, and other spin physics studies at a range of collision energies never before possible.").

156. *See* Steve Vigdor, The Case for Continuing RHIC Operations, http://www.bnl.gov/npp/docs/The%20Case%20for%20Continuing%20RHIC%20Operations%20_draft%205.pdf; *see also* BUSZA ET AL., *supra* note 5, at 21 ("At design luminosity, running for a scheduled six months per year for ten years, RHIC will produce approximately $2 \times 10^{11}$ gold–gold collisions.").

157. *See* Wolfram Fischer, *Run Overview of the Relativistic Heavy Ion Collider*, BROOKHAVEN NAT'L LABORATORY (Sept. 28, 2015), http://www.rhichome.bnl.gov/RHIC/Runs.

158. *Id.*

159. *Id.*

160. E-mail from Pete Genzer, Manager, Media & Communications Office, Brookhaven National Laboratory to Eric E. Johnson & Michael Baram (Jan. 16, 2014) (on file with author).

As a legal matter, Brookhaven appears to have little to worry about. As I discuss in the next section, a combination of administrative law and court-generated doctrines of judicial self-restraint provide a path for courts to decline to provide an on-the-merits review of the RHIC's experimental program or other particle experiments the DOE may undertake in the future.[162]

## III. THE LACK OF LEGAL REVIEWABILITY

There are two potential sources of ex ante legal constraints on federal agency science research: regulation by other agencies and review by courts. I will discuss each in turn.

### *A. Interagency Regulation*

Theoretically, there is nothing to stop Congress from structuring the administrative state such that one agency's research activity is regulated by another agency.

Following the end of World War II, Congress gave permanent form to the Manhattan Project by establishing the Atomic Energy Commission ("AEC").[163] The AEC was given general authority to issue rules and orders governing its own activities to minimize danger to persons and property.[164] In other words, it was tasked to self-regulate.[165] In 1974, the Energy Reorganization Act abolished the AEC and transferred its research functions to the newly created Energy Research and Development Administration ("ERDA"); regulatory functions were split off and placed with the newly created Nuclear Regulatory Commission ("NRC"), an independent agency headed by five commissioners.[166] Three years later, ERDA was elevated to a cabinet-level entity as the Depart-

161. *See, e.g.*, Dar et al., *supra* note 103, at 146–47 (discussing the difficulty of estimating strangelet accretion times, but providing a discussion that acknowledges accretion possibilities ranging over a term of years, centuries, or even longer, including beyond the time, billions of years from now, when the sun will engulf the Earth).

162. Michael Baram and I wrote an op-ed encouraging a federal advisory committee to consider whether an updated safety assessment for the RHIC was warranted. *See* Eric E. Johnson & Michael Baram, Op-Ed., *New U.S. Science Commission Should Look at Experiment's Risk of Destroying the Earth*, INT'L BUS. TIMES (Feb. 10, 2014), http://www.ibtimes.com/new-us-science-commission-should-look-experiments-risk-destroying-earth-1554380.

163. Atomic Energy Act of 1946, Pub. L. No. 79-585, 60 Stat. 755 (codified as amended at 42 U.S.C. §§ 2011–2296 (2012)).

164. 42 U.S.C. § 2201(i)(3).

165. *See* Barbara A. Finamore, *Regulating Hazardous and Mixed Waste at Department of Energy Nuclear Weapons Facilities: Reversing Decades of Environmental Neglect*, 9 HARV. ENVTL. L. REV. 83, 89 (1985).

166. Energy Reorganization Act of 1974, Pub. L. No. 93-438, 88 Stat. 1233 (codified as amended at 42 U.S.C. §§ 5801–5891 (2012)).

ment of Energy.[167] The NRC's establishing legislation provided that the NRC would have regulatory authority over commercial use of nuclear power and nuclear materials. With the exception of a few areas intertwined with commercial nuclear power and a limited role in high-level waste storage, the NRC was given no regulatory authority over ERDA/DOE.[168] Thus, DOE inherited the AEC's prerogative to self-regulate.

Self-regulation has worked poorly for AEC/ERDA/DOE. Working under its own auspices, the agency has, in its various historical incarnations, buried radioactive waste in cardboard boxes, allowed millions of gallons of waste to enter groundwater, and conducted medical experimentation on unwitting Americans, including injecting people with plutonium.[169]

DOE self-regulation has not gone uncriticized. In the 1990s, proposed legislation would have transitioned to a system in which DOE was regulated by other agencies, particularly the NRC.[170] Studies conducted by the NRC recommended that NRC regulate DOE facilities, concluding that "[e]xternal regulation will eliminate the inherent conflict of interest arising from self-regulation."[171] The task force also thought that external regulation would "lead to a safety culture comparable to the safety culture in the commercial industry," "enhance DOE credibility," and "enhance public confidence in DOE."[172] Ultimately, these efforts stalled, and today DOE continues to be self-regulating.[173] NASA's usage of RTGs, done in conjunction with DOE, is also a matter of self-regulation.[174]

---

167. Department of Energy Organization Act of 1977, Pub. L. No. 95-91, 91 Stat. 565 (codified as amended at 42 U.S.C. §§ 7101–7386k (2012)).

168. *See* 42 U.S.C. § 5842 (NRC regulatory authority over ERDA/DOE for demonstration liquid-metal fast breeder reactors and other demonstration nuclear reactors when operated for commercial use or for the purpose of demonstrating suitability for commercial use; depositories for commercially generated high-level radioactive waste; facilities for the fabrication of mixed plutonium-uranium oxide nuclear reactor fuel for commercial use *unless* the facility is "utilized for research, development, demonstration, testing, or analysis purposes," and long-term storage facilities for ERDA/DOE's high-level radioactive waste *unless* part of or used in research and development activities).

169. *See* ADVISORY COMMITTEE ON HUMAN RADIATION EXPERIMENTS, FINAL REPORT OF THE ADVISORY COMMITTEE ON HUMAN RADIATION EXPERIMENTS *passim* (1995) (discussing experiments conducted from 1944–1974, including plutonium injections).

170. U.S. NUCLEAR REGULATORY COMMISSION TASK FORCE ON EXTERNAL REGULATION OF DOE NUCLEAR FACILITIES, EXTERNAL REGULATION OF DEPARTMENT OF ENERGY NUCLEAR FACILITIES: A PILOT PROGRAM (NUREG-1708) 3 (1999) [*hereinafter* NUREG-1708].

171. *Id.* at ix.

172. *Id.* at 34.

173. *See* 42 U.S.C. §§ 2140, 2142, 5842.

174. *See* NASA, SAFETY OF RADIOISOTOPE POWER SYSTEMS, *supra* note 54, at 2. NASA, however, requires presidential approval before launches. *Id.*

### *B. Judicial Scrutiny*

Beyond external agency regulation, the other straight-forward mode of ex ante legal constraint on allegedly dangerous activity would be through a court-issued injunction. Unfortunately, lacking a specific invitation to the courts to do so, the law does not guarantee judicial scrutiny for federal agency projects presenting plausible catastrophic and irreversible ultrahazards.[175]

At the outset, it should be noted that a lack of judicial scrutiny is not because no one has tried to get the courts involved. Both the Cassini-plutonium and RHIC-strangelet controversies precipitated lawsuits. In 1997, a group of plaintiffs filed suit in Hawaii to stop the Cassini launch.[176] In regard to the RHIC, Walter L. Wagner filed suit in San Francisco in 1999,[177] and in New York in 2000.[178] Wagner was the *Scientific American* reader, discussed above, whose letter to the editor touched off the media attention about RHIC doomsday questions, which, in turn, spurred Brookhaven to commission its safety report.[179] So, there was litigation over both the Cassini and the RHIC disaster-risk questions. But none of it precipitated meaningful review on the issue of acceptable risk. In fact, none of the plaintiffs even achieved what could be charitably described as a moral victory.

In analyzing the availability of judicial review over these kinds of agency risks, the Cassini and RHIC litigations are instructive. The tangled nature of their pleadings, however, makes it impracticable to summarize them in chronological form. So, instead, I will analyze the availability of judicial scrutiny in the abstract, and I will reference relevant aspects of the Cassini and RHIC suits along the way.

In the analysis below, after making a preliminary observation about bare injunctions and sovereign immunity, I discuss the APA, the FTCA, NEPA, and political-question doctrine. Consideration of these aspects of the law helps frame my subsequent discussion of the problems with entrusting decisions about catastrophic and irreversible risk issues to the agencies that beget them.[180] In the service of keeping the discussion manageable and focused, I have chosen to omit consideration of additional aspects of the law that would be relevant in various circumstances to litigation over Cassini, the RHIC, or other federal science-experiment pro-grams. Those include standing, the federal-contractor defense,[181] the federal-enclave doctrine,[182] and the Price-Anderson Act.[183]

---

175. Further below, I explain why it is important for courts to provide such scrutiny. *See infra* Parts IV–V.

176. Haw. Cnty. Green Party v. Clinton, 980 F. Supp. 1160, 1163 (D. Haw. 1997).

177. Wagner v. U.S. Dep't of Energy, No. C99-2226 (N.D. Cal., compl. filed May 14, 1999). No district court opinion was published in C99-2226.

178. Wagner v. Brookhaven Sci. Assoc., LLC, No. 00-CV-1672 (E.D.N.Y., compl. filed Mar. 23, 2000).

179. *See supra* note 91 and accompanying text.

180. *See infra* Parts IV–V.

### *1. Bare Injunctions and Sovereign Immunity*

The most straightforward path to the courthouse for judicial scrutiny of allegedly risky scientific research would be an application for an equitable injunction. Courts have well-established power in equity to enjoin activities that constitute a threat to human life "even before the threat matures to result in physical injury or death."[184] When it comes to actors such as NASA or DOE, however, a bare request for an injunction will be stopped by the doctrine of sovereign immunity, which bars suits against the federal government—including its agencies—unless the government waives its immunity and consents to the suit.[185] Indeed, Wagner's San Francisco suit was a bare request for injunction based on "risk of death,"[186] and the U.S. Attorney invoked sovereign immunity as a defense.[187] Because of sovereign immunity, any plaintiff seeking ex ante judicial review of agency action will generally have to rely on some source of statutory authorization for the suit.[188]

---

181. See *In re Hanford Nuclear Reservation Litig.*, 534 F.3d 986, 1000 (9th Cir. 2008) for a general discussion of the federal-contractor defense.

182. For a general treatment, see Emily S. Miller, *The Strongest Defense You've Never Heard of: The Constitution's Federal Enclave Doctrine and Its Effect on Litigants, States, and Congress*, 29 HOFSTRA LAB. & EMP. L.J. 73 (2011).

183. Pub. L. No. 85-256, 71 Stat. 576, 42 U.S.C. § 2210, et seq. (2012).

184. Seide v. Prevost, 536 F. Supp. 1121, 1133 (S.D.N.Y. 1982); *see also* Honig v. Doe, 484 U.S. 305, 327 (1988) (holding that federal district courts have authority in equity to temporarily enjoin a child determined to be dangerous from attending a school); Harris Stanley Coal & Land Co. v. Chesapeake & Ohio Ry. Co., 154 F.2d. 450, 453–54 (6th Cir. 1946) (holding that district court abused its discretion by failing to issue an injunction to restrain a coal-mining company from underground "pillar pulling" operations that posed a risk to railroad passengers travelling on track on the surface); Shimp v. N.J. Bell Tel. Co., 368 A.2d 408, 411–12 (N.J. Super. Ct. Ch. Div. 1976) (issuing an injunction prohibiting employees from smoking near a person whose health issues made the tobacco smoke hazardous to her).

185. *See* United States v. Shaw, 309 U.S. 495, 500–01 (1940), *cited in* Federal Defendant U.S. Dep't of Energy's Notice of Motion and Motion to Dismiss or, in the Alt., for Summary Judgment at 4, Wagner v. U.S. Dep't of Energy, No. C99-2226 (N.D. Cal., May 5, 2000) (citation given as "309 U.S. 495 (1959)").

186. Complaint and TRO Request at 3, Wagner v. U.S. Dep't of Energy, No. C99-2226 (N.D. Cal., May 14, 1999).

187. Defendant U.S. Dep't of Energy's Motion to Dismiss, *supra* note 185, at 4. The court did not issue any rulings on the grounds of sovereign immunity.

188. Note that where a suit against the United States finds footing in the Constitution, such statutory authorization may be unnecessary. *See, e.g.*, Bivens v. Six Unknown Named Agents of the Fed. Bureau of Narcotics, 403 U.S. 388, 404 (1971) (Harlan, J., concurring) (noting "the presumed availability of federal equitable relief against threatened invasions of constitutional interests"); *see also* Anya Bernstein, *Congressional Will and the Role of the Executive in* Bivens *Actions: What Is Special About Special Factors?*, 45 IND. L. REV. 719, 725–26 (2012) (discussing historical availability of injunctions against threatened government incursions on individual rights). There is, however, a dearth of precedent for using due process guarantees as a way to challenge low-probability risks such as those under discussion.

### 2. *The Administrative Procedure Act*

The most natural place to look for a statutory footing in suing an agency is the Administrative Procedure Act ("APA"), the statute setting out the comprehensive legal framework that governs agency procedure.[189] The APA provides for judicial review of agency actions in various circumstances, and, where it does so, sovereign immunity is not an issue.[190]

Does the APA provide a way for plaintiffs to challenge agency science programs that allegedly constitute catastrophic and irreversible ultrahazards?

The short answer is yes—at least for a court undaunted by the technical and political difficulties posed by such a case. The text of the statute and various cases interpreting it provide a pathway for judicial review. Yet there is also a lack of clarity in the law, and one can find plenty of footholds for arguments against judicial review. Moreover, recent cases imply that the difficulty of judicial review militates against its appropriateness,[191] and when it comes to leading-edge science experiments rife with uncertainty, difficulty can be found in every direction.[192] Add to that what Emily Hammond Meazell has called "a natural judicial tendency to avoid any deep confrontations with science,"[193] and ultimately the APA does not, as a practical matter, guarantee judicial review for agency decisions about catastrophic and irreversible ultrahazards stemming from agency research programs.

The long answer to whether the APA provides a basis for judicial review is more complicated and more nuanced. It is helpful to think about it from two different frames of reference—the long view, looking at the statute as it is perceived in its entirety, and the close-up view, looking at particular provisions.

#### a. The APA's Gestalt

At the broadest level of generality, the APA appears as something of a mismatch for plaintiffs challenging government science experiments. The problem might be described as one of gestalt.[194] In terms of its overall design, how it functions in practice, and how it is regarded by courts and commentators, the APA is a kind of constitutional document for the administrative "branch" of the federal government.[195] And in the imagination of the APA, government agencies *govern*. In other words, the APA is primarily concerned with how agencies use their coercive authority in ways that require a balancing of private interests against the broader public good.[196] Think of agency decisions such as whether an oil refinery can emit certain pollutants, whether a given individual will be allowed to pilot an aircraft, or whether a cosmetics company can use a certain colorant in eyeshadow.

---

189. 5 U.S.C. §§ 500–706 (2012).

190. *See id.* § 704 ("Agency action made reviewable by statute and final agency action for which there is no other adequate remedy in a court are subject to judicial review.").

191. *See* Brief for Administrative Law Professors as Amici Curiae in Support of Petitioner at 17–18, Ochoa v. Holder, 131 S. Ct. 3058 (2011) (No. 10-920).

192. In Part V, *infra*, I explain why this difficulty ought not be a barrier to judicial review.

193. Emily Hammond Meazell, *Super Deference, the Science Obsession, and Judicial Review As Translation of Agency Science*, 109 MICH. L. REV. 733, 734 (2011).

194. The word *gestalt* refers to "an organized whole that is perceived as more than the sum of its parts." CONCISE OXFORD AMERICAN DICTIONARY 376 (2006).

In imagining agencies as governing institutions, the APA recognizes that agencies engage in quasi-legislative and quasi-judicial behavior—rulemaking and adjudication, respectively.[197] Based on the statutory attention lavished on these subjects, it is clear that the APA sees these activities as the two most important modes of agency work. Recognizing their importance and potential for mischief, the APA restrains agency conduct in these arenas and, in so doing, provides a measure of protection for the citizens whose rights may be affected. Specifically, the APA attempts to ensure fairness and rationality in agency rulemaking and adjudication through such means as requiring notice, providing rights to be heard, and recognizing mechanisms for judicial review.[198]

---

195. *See, e.g.*, Reginald Parker, *The Administrative Procedure Act: A Study in Overestimation*, 60 YALE L.J. 581, 583 (1951) ("The basic purpose of the APA was obviously the wish to bring about, some how, a curb of the administrative branch of our government—to see to it that the governors shall be governed and the regulators shall be regulated.") (internal quotation marks omitted); Antonin Scalia, *Vermont Yankee: The APA, the D.C. Circuit, and the Supreme Court*, 1978 SUP. CT. REV. 345, 363 (1978) ("[T]he Supreme Court regarded the APA as a sort of superstatute, or subconstitution, in the field of administrative process: a basic framework that was not lightly to be supplanted or embellished . . . ."); Alan B. Morrison, *The Administrative Procedure Act: A Living and Responsive Law*, 72 VA. L. REV. 253, 253 (1986) ("[T]he APA is more like a constitution than a statute."); Note, *Deportation and Exclusion: A Continuing Dialogue Between Congress and the Courts*, 71 YALE L.J. 760, 791 (1962) (noting that the APA "embodies a quasi-constitutional guarantee of fairness which attempts to take account of varying needs in different applications of the administrative process"); *see also* Kathryn E. Kovacs, *Leveling the Deference Playing Field*, 90 OR. L. REV. 583, 607 n.156 (2011) (collecting cites of commentators agreeing that "the APA has taken on quasi-constitutional status").

196. *See, e.g.*, 5 U.S.C. § 551 (defining "agency proceeding" as meaning rulemaking, adjudication, or licensing, and "agency action" as "includ[ing] the whole or a part of an agency rule, order, license, sanction, relief, or the equivalent or denial thereof, or failure to act").

197. *See id.*; *see also, e.g.*, Portland Audubon Soc. v. Endangered Species Comm., 984 F.2d 1534, 1540 (9th Cir.1993) ("Where an agency's task is to adjudicate disputed facts in particular cases, an administrative determination is quasi-judicial. By contrast, rulemaking concerns policy judgments to be applied generally in cases that may arise in the future[.]") (citations and internal quotation marks omitted); Sokaogon Chippewa Cmty. (Mole Lake Band of Lake Superior Chippewa) v. Babbitt, 929 F. Supp. 1165, 1174 (W.D. Wis. 1996) on reconsideration in part, 961 F. Supp. 1276 (W.D. Wis. 1997) ("There are two basic types of agency actions: rulemaking (quasi-legislative) and adjudication (quasi-adjudicative). The distinctions between the two can be murky; however, in general, rulemaking comprises the formulation of policies of widespread application while adjudication covers the resolution of disputes between specific parties.") (citation omitted).

198. *See, e.g.*, 5 U.S.C. § 553 (requiring notice of proposed rulemaking); *id.* § 555(b) (right to appear in proceeding); *id.* § 702 (right of judicial review).

In addition to being the major preoccupations of the APA's text and structure, the functions of adjudication and rulemaking are the main focus of courts and commentators seeking to clarify administrative law doctrine. Commentators have also weighed in on agencies' enforcement functions and prosecutorial conduct,[199] but even these activities are like rulemaking and adjudication in that they involve governing-type tasks.

In contrast to its intense focus on the administrative governing process, the APA concerns itself only lightly with what I will call *non-governing* activities, that is, activities that involve no use of coercive authority to curb the conduct of private parties. Non-governing activities are ones that private parties could theoretically engage in.

Scientific research is a non-governing activity undertaken by agencies. A private university, a privately held business, or even an individual can conduct scientific research.[200] Studying quark-gluon plasma, as the RHIC does, or exploring Saturn, as Cassini was built to do, are good examples.[201] The reasons it falls to federal agencies to pursue such projects is mostly a function of their expense. It takes the pocketbook of a sovereign government to build a multi-billion-dollar machine for the purpose of satisfying scientific curiosity.[202] Yet the activity is nothing like rulemaking or adjudication, and thus it is not the kind of agency function clearly contemplated by the APA's architecture.

To indulge in a bit of anthropomorphizing, when the APA says "agency," it seems to be thinking of OSHA, the EPA, or the FCC—not NASA or Brookhaven. None of this would be a problem if non-governing agency action was incapable of treading on the rights and interests of citizens. And under ordinary circumstances, private-actor-type conduct by agencies does not threaten the interests of citizens, unless one counts generic objections that agencies are wasting taxpayers' money. This is what makes the specter of laboratory disasters interesting: They potentially pose a direct physical threat to private persons and property.

Given the historical timeline, it should not be surprising that Congress was not thinking of "big science" when it created the APA. Con-gress passed the APA on June 11, 1946.[203] In doing so, it was looking backward to the great expansion of the regulatory state that occurred during the New Deal. The explosive growth of federal science initiatives took place after the APA's passage. In fact, two of the most important science agencies were created just weeks after Congress's work on the APA was finished: The Centers for Disease Control and Prevention ("CDC") was established on July 1, 1946,[204] and the Atomic Energy Commission, forerunner to DOE, was established by legislation passed on August 1, 1946.[205] NASA came much later, established by the National Aeronautics and Space Act on July 29, 1958.[206]

---

199. *See, e.g.*, Richard M. Thomas, *Prosecutorial Discretion and Agency Self-Regulation:* CNI v. Young *and the Aflatoxin Dance*, 44 ADMIN. L. REV. 131 (1992) (arguing that D.C. Circuit jurisprudence is at odds with Supreme Court precedent on the issue of agency deference in prosecutorial discretion).

200. The cleavage between governing and nongoverning agency action may not always be so clear. Some research agendas are intimately tied to quintessentially governmental objectives—an obvious example being nuclear weapons research. Such laboratory research might be thought of as quasi-governing. Many other scientific research agendas, however, bear no special relationship to government objectives. Both Cassini and the RHIC fall into this later category.

201. *See supra* Part II.B–C.

202. *See, e.g.*, *Our History,* BROOKHAVEN NATIONAL LABORATORY, https://www.bnl.gov/about/history/ ("Brookhaven was originally conceived, in part, to establish a national laboratory in the Northeastern United States to design, construct and operate large scientific machines that individual institutions could not afford to develop on their own.") (last visited Jan. 24, 2016).

The bottom line is that the APA, as a whole, appears to be talking past the kinds of agency actions that may create catastrophic and irreversible ultrahazards.

b. The APA's Dueling Provisions on Judicial Review

The general awkwardness of the APA in the context of agency science risk is not alleviated by focusing on its specific provisions.

Section 701(a)(2) of the APA states that its judicial review provisions apply "except to the extent that . . . agency action is committed to agency discretion by law."[207] By its letter, this provision might seem to make a vast swath of agency actions unreviewable—the conducting of science experiments among them. But the apparent prohibition of § 701(a)(2) is not the end of the story. Substantial legislative history indicates that § 701(a)(2)'s apparent bar on judicial review was meant only to express "the truism that, when an agency has been granted broad discretionary powers, its actions are likely to be lawful and therefore immune from judicial reversal."[208] What is more, a few pages after § 701(a)(2), the APA seems to make a broad endorsement of widely available judicial review. Section 706(2)(A) provides, "The reviewing court shall hold unlawful and set aside agency action . . . found to be arbitrary, capricious, [or] an abuse of discretion."[209] The effort to make sense of all this impli-

203. Pub. L. No. 79-404, 60 Stat. 237.

204. *Our History–Our Story*, CTRS. FOR DISEASE CONTROL & PREVENTION, http://www.cdc.gov/about/history/ourstory.htm (last visited Jan. 24, 2016).

205. Atomic Energy Act of 1946, Pub. L. No. 79-585, 60 Stat. 755; *see also* ALICE BUCK, U.S. DEP'T OF ENERGY, THE ATOMIC ENERGY COMMISSION 1, 19 (1983), *available at* http://energy.gov/sites/prod/files/AEC%20History.pdf.

206. Note that the National Advisory Committee for Aeronautics ("NACA"), a federal research agency that was a forerunner to NASA, was established March 3, 1915. In its infancy, however, NACA was very small—comprising a staff of 15. *See* Elizabeth Suckow, *Overview*, NASA, http://history.nasa.gov/naca/overview.html (last updated Apr. 23, 2009).

207. 5 U.S.C. § 701(a)(2) (2012).

208. Ronald M. Levin, *Understanding Unreviewability in Administrative Law*, 74 MINN. L. REV. 689, 696–97 (1990).

209. *Id.* § 706(2)(A).

cates a running conversation in the administrative law literature—one described as an "epic scholarly debate."[210]

The controversy has not been resolved by the courts. In fact, cases point in different directions. The landmark opinion in *Citizens to Preserve Overton Park, Inc. v. Volpe* characterized § 701(a)(2) as "a very narrow exception" to the usual availability of judicial review, holding that it is "applicable in those rare instances where statutes are drawn in such broad terms that in a given case there is no law to apply."[211] In that case, the Court upheld the availability of judicial review over the Department of Transportation's decision to build Interstate 40 through a public park in Memphis, Tennessee.[212] But later cases have moved away from *Overton Park*'s liberal attitude toward judicial review, directing courts to consider the level of difficulty a court would face in providing review.[213] As summed up by a group of administrative law scholars, "In simplest terms, there is no clear, coherent framework for evaluating claims of unreviewability under Section 701(a)(2). As a result, there is a state of confusion in the lower courts as to when Section 701(a)(2) applies."[214]

Meanwhile, it is unclear to what extent § 706(2)(A) gives plaintiffs a foothold for review of non-rulemaking, non-adjudicatory activity. An argument for the availability of review via § 706(2)(A) can be made from what Ronald M. Levin has called a "pure abuse of discretion" theory.[215] Broadly construed, this theory of judicial review would allow virtually any agency action to be challenged in the courts. Levin writes that, in pursuing a pure abuse of discretion theory, "[a] plaintiff might claim, for example, that the agency misunderstood the facts, that it departed from its precedents without a good reason, that it did not reason in a minimally plausible fashion, or that it made an unconscionable value judgment."[216]

Such a broad view of judicial review is not precluded by case law, but courts may be reluctant to embrace it, at least absent a strong rationale for why it is warranted.[217] Regardless, there is a regrettable lack of

210. *See* Brief for Administrative Law Professors, *supra* note 191, at 10.

211. 401 U.S. 402, 410 (1971) (internal quotation marks omitted).

212. *Id.* at 406–07, 421.

213. *See, e.g.*, ICC v. Bhd. of Locomotive Eng'rs, 482 U.S. 270, 282 (1987) (holding an agency's refusal to reconsider for material error to be nonreviewable, referring to "the impossibility of devising an adequate standard of review for such agency action"). With regard to the potential difficulty of judicial review in cases of agency-created science-experiment risk, I argue, further below, that courts are actually well-suited to the task of providing review of such agency decisions, even where the scientific subject matter is arcane. *See infra* Part V.

214. Brief for Administrative Law Professors, *supra* note 191, at 21.

215. Levin, *supra* note 208, at 708.

216. *Id.*

217. Parts IV and V provide an argument for why such review is needful in low-probability/high-harm risk scenarios implicated by agency science experiments, and, by extension, other non-governing agency actions.

clarity on the issue. Legal scholar Gordon G. Young has called § 706 "a disorderly mess of ambiguous and overlapping standards."[218] At the end of the day, the practical availability of judicial review over discretionary-type agency decisions may be regrettably low. Legal scholar Mariano-Florentino Cuéllar has observed that the role of the courts in reviewing decisions within executive discretion "is often so circumscribed or deferential that in some domains the probability of uncovering problems through such review almost certainly falls close to zero."[219]

### 3. *The Federal Tort Claims Act*

The APA notwithstanding, there is, within federal statutory law, a crystal-clear recognition that agencies do more than govern and that these non-governing activities have the potential to cause physical injury: the Federal Tort Claims Act ("FTCA"), which allows tort lawsuits against the United States.[220] But rather than submit the federal government to the tort scheme *en masse*, the FTCA provides only for a carefully controlled window of liability. Among its many procedural and substantive provisions are two limitations that are important for catastrophic and irreversible ultrahazards. First, the FTCA's only remedy is compensatory damages—the act does not authorize injunctive relief or otherwise provide a vehicle for ex ante review of allegedly dangerous activities.[221] Second, FTCA liability is limited by a powerful discretionary-function exception.[222] This provision is designed "to prevent judicial 'second-guessing' of . . . administrative decisions grounded in social, economic, and political policy through the medium of an action in tort."[223] As a result of these limitations, the FTCA prevents tort law from being used to challenge agency decisions on the acceptability of risk from government laboratory activities.

The FTCA fits the same historical pattern as the APA. Like the APA, the FTCA arose before the "big science" era.[224] To anthropomorphize: The FTCA sees the world of government risk as comprising postal trucks running red lights—not mass death spawned by an out-of-control multi-billion-dollar science project.

---

218. Gordon G. Young, *Judicial Review of Informal Agency Action on the Fiftieth Anniversary of the APA: The Alleged Demise and Actual Status of* Overton Park*'s Requirement of Judicial Review "On the Record"*, 10 ADMIN. L.J. AM. U. 179, 181 (1996).

219. Mariano-Florentino Cuéllar, *Auditing Executive Discretion*, 82 NOTRE DAME L. REV. 227, 227 (2006).

220. *See supra* note 10.

221. *See, e.g.*, Moon v. Takisaki, 501 F.2d 389, 390 (9th Cir. 1974) (holding that FTCA does not submit the federal government to injunctive relief).

222. 28 U.S.C. § 2680(a) (2012) (FTCA liability "shall not apply to . . . the failure to exercise or perform a discretionary function or duty on the part of a federal agency or an employee of the Government, whether or not the discretion involved be abused").

223. United States v. S.A. Empresa de Viacao Aerea Rio Grandense (Varig Airlines), 467 U.S. 797, 814 (1984).

224. *See supra* note 203 and accompanying text.

Although it can be argued that the APA provides, by its letter, an avenue for the review of potentially risky science experiments, there is not even a theoretical opening under the FTCA to do so. The closure of the FTCA to such suits is clear.

An apposite case is *Konizeski v. Livermore Labs (In re Consolidated U.S. Atmospheric Testing Litigation)*, decided in 1987, in which the Ninth Circuit held that the FTCA did not allow a negligence claim for radiation injuries caused by the atmospheric detonation of nuclear weapons.[225] Plaintiffs argued that "the exercise of . . . 'scientific or professional judgment' in implementing an overall decision to conduct the atomic weapons testing program" was not covered by the discretionary-function exception.[226] The court disagreed, holding that safety decisions were part of the policy decisions required for the program.[227] In elaborating, the court expressed a veneration for agency expertise—as well as a strong perception of its own incompetence in the area of nuclear-experimentation risks:

> A court would be ill-equipped to evaluate the judgments concerning safety made by those officials based on the exigencies of the moment. Any attempt to do so would, moreover, require a comprehensive reexamination of the conduct of the tests and the decisions made during their course which would itself defeat the purpose of the exception. The consequences of such a reexamination would be to hamper the government in its future conduct of weapons tests and similar operations affecting the national security.[228]

In the years since *Konizeski,* courts have embraced the discretionary-function exception in an ever expanding variety of situations. In *Kohl v. United States*, Debra R. Kohl, a bomb technician with the Metropolitan Nashville Police Department, was injured during a training program conducted by Oak Ridge National Laboratory.[229] A government employee used a winch to pull on the door of an explosion-damaged car while Kohl was leaning into the vehicle from the other side.[230] The winched part suddenly came loose, causing the vehicle to crash into Kohl's head.[231] She was left with "post-concussive syndrome with persistent headaches

225. 820 F.2d 982, 996–99 (9th Cir. 1987).
226. *Id.* at 993.
227. *Id.* at 995 ("[W]here there is room for policy judgment and decision there is discretion.") (quoting Dalehite v. United States, 346 U.S. 15, 36 (1953)).
228. 820 F.2d at 995.
229. Kohl v. United States, 699 F.3d 935, 938–39 (6th Cir. 2012).
230. *Id.* at 938–39.
231. *Id.* at 939.

and cognitive changes."[232] Although calling it a "close case," the *Kohl* court held the discretionary-function exception applied. It explained:

> The decision to use a winch was part of the decisionmaking involved in deciding how best to conduct the post-blast investigation. . . . The planning and execution of the research experiment is susceptible to policy analysis, including judgments about how to respond to hazards, what level of safety precautions to take, and how best to execute the experiment in a way that balanced the safety needs of the personnel and the need to gather evidence from the vehicles.[233]

The *Kohl* opinion is certainly open to criticism. Indeed, a dissent in *Kohl* called the majority's theory "incoherent."[234] Nonetheless, given precedents such as *Kohl,* plaintiffs' prospects for challenging science-experiment design under an FTCA framework would be very dim—even if the FTCA were amended to allow for injunctive relief. If carelessly using a winch to yank on a car with a person leaning inside is an FTCA-exempted discretionary decision, then the decision to launch a rocket or operate a particle accelerator would almost certainly fall within the discretionary-function exception, as it is currently construed.

### 4. *The National Environmental Policy Act*

Plaintiffs trying to halt federal science endeavors on the basis of alleged catastrophe risk have generally resorted to the National Environmental Policy Act of 1969 ("NEPA").[235] As the nation's premiere law on the environmental consequences of federal action, NEPA requires federal agencies to prepare and file an environmental impact statement ("EIS") before proceeding with any major action that may significantly affect the quality of the environment.[236] The EIS must discuss adverse environmental effects and explain what alternatives exist to the proposed action.[237] A person believing that an agency failed to live up to the EIS requirement of NEPA can sue. The court's job with such a lawsuit is to read the EIS to see if the agency took a "hard look" at the environmental consequences of its proposed action.[238]

---

232. *Id.*

233. *Id.* at 943.

234. *Id.* at 946 (Merritt, J., dissenting). The dissent explained, "The problem with formulating a standard or principle [as the court does] is that almost every act by government or private agent in the scope of employment would 'challenge a policy' if it is for the purpose of carrying out some government or private interest, policy or plan." *Id.*

235. 42 U.S.C. §§ 4321–4347 (2012).

236. *Id.* § 4332(2)(C).

237. *Id.*

238. Friends of the Payette v. Horseshoe Bend Hydroelectric Co., 988 F.2d 989, 993 (9th Cir. 1993), *cited* by Haw. Cnty. Green Party v. Clinton, 980 F. Supp. 1160, 1166 (D. Haw. 1997).

NEPA was the sole claim in Wagner's New York lawsuit against the RHIC. That suit, however, never reached NEPA issues on the merits because it was dismissed on procedural grounds related to Wagner having previously filed a similar lawsuit in San Francisco.[239]

A lawsuit that was focused on NEPA and reached the issue of NEPA compliance was *Hawai'i County Green Party v. Clinton,* the suit to stop the Cassini launch.[240] There was no question as to whether NEPA applied to NASA's actions.[241] The problem for the plaintiffs was that NASA handily showed that it had done all it was required to do under the law.[242] The district court found that the government "demonstrated to the court that environmental consequences, potential accident scenarios and potential health risks were carefully and adequately considered in the research for the Cassini Mission."[243] Whether Cassini was unreasonably dangerous was irrelevant. As the opinion explained, "The court is not entitled to second guess the expertise of the federal agency, it can only inquire into whether all required considerations were made."[244]

The Cassini litigation illustrates that the EIS requirement, by itself, can do nothing to halt damaging actions to the environment. At bottom, NEPA does not provide a path for genuine judicial review of agency action on the merits.

That is not to say that NEPA does not have beneficial effects for the environment. NEPA can create better environmental outcomes in at least two ways. Neither, however, is likely to be an effective constraint when it comes to low-probability catastrophic risk of the type discussed in this Article.

First, NEPA can create positive environmental outcomes because of the exercise of preparing the EIS. And the interaction NEPA fosters with experts at the Environmental Protection Agency may lead to productive soul-searching within an agency, prompting the agency to modify its plans accordingly. An agency might, for example, change construction blueprints, make alternative arrangements for handling hazardous materials, or the like. But in a case like Cassini, the full environmental risk, although perhaps capable of some amelioration, is not easily detachable from the project itself.[245] And for the RHIC, the risk would seem to be

239. Wagner v. Brookhaven Sci. Assoc., LLC, No. 00-CV-1672 (E.D.N.Y., compl. filed Mar. 23, 2000), Order, at 2, May 31, 2000.

240. 980 F. Supp. 1160 (D. Haw. 1997).

241. *Id.* at 1167–68, 1170–71.

242. *Id.* at 1168, 1171.

243. *Id.* at 1167.

244. *Id.* at 1167–68.

245. *See, e.g*, NASA, VVEJGA TRAJECTORY at 2 (2011), http://saturn.jpl.nasa.gov/files/trajectory.pdf (discussing why the Earth-swingby trajectory was used to reach Saturn, since current rockets are not large enough to create the speed needed for a direct trajectory).

entirely bound up with the endeavor.[246] If shuttered, such a project would take with it billions of dollars in sunk costs and upend countless careers.

Second, assuming the agency preparing the EIS is not disposed to change its mind about its contemplated action, the EIS requirement can still act as a constraint on bad agency choices by enabling the public to engage with politicians and bureaucrats on the ultimate issue of whether the action is worthwhile considering the harm it might do. The EIS, when made available to the public, "foster[s] . . . informed public participation."[247] In other words, NEPA serves an information-forcing function. Yet as I explain further below, however, there is strong reason to believe that political control over agencies will be insufficient to adequately protect the public against Cassini/RHIC-type risks.[248]

### 5. *Political-Question Doctrine*

The interface of agency-science risk and politics brings up another feature of American jurisprudence that one might expect to surface in litigation of this type: political-question doctrine. As enunciated by the Supreme Court in *Baker v. Carr*, whether an issue before the courts is nonjusticiable as a political question "is primarily a function of the separation of powers."[249] In deciding whether something is a political question, "the appropriateness under our system of government of attributing finality to the action of the political departments and also the lack of satisfactory criteria for a judicial determination are dominant considerations."[250]

Although political-question doctrine was not invoked in the Cassini or RHIC litigations, one imagines it might be in a similar suit in the future. A court came close to calling on the doctrine in 2008, when Walter Wagner and one other person filed a lawsuit challenging DOE's participation in a particle accelerator project in Europe—primarily on the basis that it could generate a planet-destroying black hole.[251] In that case, *San-*

246. Some sources do suggest possible means of risk mitigation for the RHIC. *See, e.g.*, Kent, *Critical Look*, *supra* note 3, at 10 ("[H]ad the problem of reducing the risk bounds been taken seriously, further theoretical research, perhaps combined with a tentative experimental programme aimed at carefully testing our understanding of the new physics involved before running the full experiment, could almost certainly have reduced the bounds very significantly."). But the technical feasibility of risk mitigation is much different than its budgetary or political feasibility. Acknowledging a need to delay or modify a project like the RHIC on account of a risk of global destruction could well spell the project's fiscal and political death.

247. California v. Block, 690 F.2d 753, 761 (9th Cir. 1982), *cited in* Haw. Cnty. Green Party v. Clinton, 980 F.Supp. 1160, 1167 (D. Haw. 1997).

248. *See infra* Part IV.E.

249. Baker v. Carr, 369 U.S. 186, 210 (1962).

250. *Id.* (quoting Coleman v. Miller, 307 U.S. 433, 454–55 (1939)).

251. *See* Sancho v. U.S. Dep't of Energy, 578 F. Supp. 2d 1258, 1259 (D. Haw. 2008). For detailed discussion of the black-hole scenario and how a court ought to deal with such a challenge, see Johnson, *The Black Hole Case: The Injunction Against the End of the World*, *supra* note 94; *see also* Eric E.

*cho v. Department of Energy*, the court said, "[t]he political process, and not NEPA, provides the appropriate forum in which to air policy disagreements."[252]

The *Sancho* court's nod toward the political process, however, raises a conundrum in a NEPA case. Since NEPA serves an information-forcing function to support an engagement of the political process, declining to bring NEPA to bear on a dispute because of a preference for the political process seems to put the cart before the horse. Moreover, political-question doctrine seems inapposite for the same reason I suggest NEPA itself is problematic in this context: The political process seems unlikely to reach efficient and rational outcomes on such matters.[253]

## IV. Inadequacies of Entrusting Decisions to Agencies

The absence of interagency regulation and the existence of barriers to thorough judicial review means that questions of catastrophic risk may be left to the laboratory-operating agencies themselves. In this part, I argue that this mode of governance is undesirable because we can expect it to produce objectively flawed decision-making: For a variety of reasons, we should not trust agencies to adequately protect the public interest when considering potential catastrophic risks posed by their own scientific research.

To be clear, I am not making a general indictment of the administrative state. Given the kinds of procedural safeguards provided by the APA, agencies can do a good job of protecting the public interest when making regulatory-type decisions affecting the conduct of third parties. My point, instead, is that there is a problem in circumstances of self-interest: If an agency's own scientific program presents a plausible risk of catastrophe, then the agency itself should not be trusted to make the decision about whether that risk is acceptable.

Some will find the point I am trying to make obvious. Others will find it dubious. How any given person sees the issue may depend on what one brings to the table in terms of underlying assumptions about what it is that scientists are doing when they assess the riskiness of their experiments.

One view is that when science-conducting agencies do risk assessment, they are acting as decision-makers charged with upholding the public interest. This might be called the "process-centered view." Seen

---

Johnson, *CERN on Trial: Could a Lawsuit Shut the LHC Down?*, New Scientist, Feb. 17, 2010, at 24–25.

252. 578 F. Supp. 2d 1258, 1269 (D. Haw. 2008) (quoting Metro. Edison Co. v. People Against Nuclear Energy, 460 U.S. 766, 777 (1983)).

253. *See infra* Part IV.E.

from this perspective, my contention may seem straightforward: When it comes to decision-making in the public interest, self-dealing always raises red flags. The law recognizes this generally. A judge who is part owner of a business entity, for instance, cannot preside over litigation in which that business entity is party; nor can a judge preside over a matter in which the judge has personal knowledge of disputed facts.[254] Judges in such positions must recuse themselves.[255] Following that logic, agencies should not be the referees of whether their own conduct is impermissibly risky.

Another view is that when science-conducting agencies do risk assessment, they are doing science. This could be called the "science-centered view." For example, the Busza and Dar papers[256] regarding the potential risks of heavy-ion collisions, evince this science-centered view. From this perspective, doing risk assessment is like doing primary scientific research—it involves making use of empirical data and applying deductive and inductive reasoning. If one sees risk-assessment work to be nothing more than the practicing of science, the self-recusing judge example may appear inapposite. One might reason that there is no concern for bias or conflicts of interest among scientific experts so long as they are making judgments according to scientific methods.

To cover all bases, I am aiming my argument here at an audience with a science-centered view. Thus, I take seriously the supposition that scientific methods reach above the ordinary problems that come with self-dealing.

### *A. Moral Hazard and Agency Self-Interest*

As an initial step, I think it is helpful to frame the conflict-of-interest problem in economic terms, using the concept of *moral hazard*.[257] Moral hazard describes a situation in which "someone other than a buyer pays for the buyer's purchases."[258] The fact that the buyer does not incur the full cost causes the buyer to purchase irresponsibly.[259] If you give your niece and nephew a $100 bill and ask them to go to the store to buy a light snack, theory says you can expect that they will not carefully consider prices.

---

254. *See, e.g.*, CODE OF CONDUCT FOR UNITED STATES JUDGES Canon 3(C) (1973) ("A judge shall disqualify himself or herself in a proceeding in which the judge's impartiality might reasonably be questioned, including but not limited to instances in which . . . the judge has a personal bias or prejudice concerning a party, or personal knowledge of disputed evidentiary facts concerning the proceeding . . . [or] the judge . . . has a financial interest in the subject matter in controversy or in a party to the proceeding, or any other interest that could be affected substantially by the outcome of the proceeding . . . ").

255. *See id.*

256. BUSZA ET AL., *supra* note 5; Dar et al., *supra* note 103. I discuss these papers in Part II.C., *supra*.

257. The term "moral hazard" is potentially confusing, since it has nothing to do with a "hazard" in the sense of physical peril. Moreover, its connection to morals is tenuous at best.

258. Stephen Breyer, *Analyzing Regulatory Failure: Mismatches, Less Restrictive Alternatives, and Reform*, 92 HARV. L. REV. 549, 557 (1979), *cited in* KENNETH F. WARREN, ADMINISTRATIVE LAW IN THE POLITICAL SYSTEM 81 (5th ed., 2011).

259. *Id.*

In standard neoclassical economics, a free market generally produces efficient or welfare-maximizing outcomes where consumers make purchasing decisions according to their own self-interest. In other words, the free market avoids waste because the amount of value consumers receive from their purchases corresponds to their willingness to pay. For example, the reason consumers do not buy more milk than they need is that they will only incur the cost of buying the milk if the value they receive is equal to or more than the price they pay. Moral hazard describes a situation where the linkage between value received and willingness to pay is broken. That is, if someone else is picking up the tab, then consumers will, according to theory, over-consume. Why not take home an extra gallon of milk—even if it will likely spoil before you drink it—as long as you are not paying for it? Because moral-hazard situations can produce economic inefficiency, moral hazard is often considered an economically sound reason for imposing regulation on markets.[260]

Moral hazard is not limited to purchase prices. In a broader sense, moral hazard is implicated whenever harm results to persons outside the economic transaction, something economists call negative externalities. Pollution is a traditional example of a negative externality. Leaded gasoline causes pollution. Yet leaded gas is cheaper to manufacture, and thus its price is lower than unleaded gas. And leaded gas performs well in internal-combustion engines. Left to their own devices, consumers and oil companies may be happy to enter transactions for the sale of leaded gasoline. Self-interested consumers and companies both benefit. And in a free market, the retail price of leaded gas will reflect how expensive it is to produce the gasoline. But the "true cost"—which includes environmental damage—is greater. Regulation, by banning the sale of leaded gasoline, constrains the free market and prevents society as a whole from absorbing the negative externalities.

Although purchasing decisions are the chief example of moral hazard, the concept is potentially applicable to any decision—whether or not denominated in money. Any time one person stands to receive a benefit and the cost will be absorbed by someone else, the moral hazard concept applies.

A simple and colorful example of moral hazard in an agency-science context can be found in the autobiography of NASA space shuttle astronaut Mike Mullane.[261] When Mullane first went into space, he was lucky enough to draw a seat on the shuttle's upper deck, which meant that on his return to Earth, he could see out the windows and enjoy spectacular views of the orbiter's fiery re-entry.[262] On his second mission, however, Mullane was disappointed to be assigned a seat on *Atlantis*'s windowless lower-deck.[263] When it came time for de-orbiting procedures, Mullane asked the shuttle's commander if he could hang out for a while on the upper deck to shoot video footage—with the understanding that he would get to his lower-level seat before the G-forces became too strong.[264] The commander obligingly went off-checklist, bending the rules to let Mullane stay.[265] Mullane enjoyed the sights as the light show came on. Yet despite Mullane's prior assurances, when the commander announced "Gs starting to build," Mullane remained on the upper deck.[266] He stayed even though his muscles, acclimated to weightlessness, were already struggling to hold up his body.[267] Watching the plasma streamers out the windows, Mullane noticed that they were brighter than those he saw on his first mission; that, in turn, caused Mullane to wonder if *Atlantis* might be burning up.[268] Yet Mullane still didn't go below deck to his seat. [269] Instead, he remained on the upper deck as *Atlantis* descended into the atmosphere, wowed as "clouds appeared to skim by at science-fiction speeds."[270] Finally, unable to bear the G-forces, Mullane collapsed to the floor.[271] "It was beyond time to get to my seat," he recalls.[272] "I was stuck on the flight deck, its steel floor now my seat, a situation I didn't altogether regret. I hated the thought of being downstairs. I would have been staring at a wall of lockers."[273]

---

260. *Id.*
261. MIKE MULLANE, RIDING ROCKETS: THE OUTRAGEOUS TALES OF A SPACE SHUTTLE ASTRONAUT (2006).

What is especially remarkable about Mullane's failure to return to his seat is that his assigned position was the one closest to the escape hatch, and it was Mullane's job to be ready in an emergency to help the crew bailout by jettisoning the hatch and deploying the slide pole.[274]

The situation represents a moral hazard because Mullane gained a benefit—watching the spectacle of atmospheric re-entry—but he did not pay the full cost—an elevated risk of death for each of the other six astronauts on board.

Mullane's example is a particularly helpful one when considering risks created by agency activities, because it illustrates that moral hazard

262. *Id.* at 186–87 (discussing mission STS-41-D on *Discovery*).
263. *See id.* at 285, 287–88 (discussing mission STS-27).
264. *Id.* at 285.
265. *See id.*
266. *Id.* at 286–87.
267. *Id.*
268. *See id.*
269. *Id.*
270. *Id.* at 287.
271. *Id.*
272. *Id.*
273. *Id.*
274. *Id.* at 288.

applies even when the decision-maker absorbs some, but not all, of the cost. If the RHIC imposes a slight risk of destroying the planet, then the nuclear scientists in charge of the RHIC program are subjecting themselves to that risk along with everyone else. The same is true of a spacecraft that might cause a risk of additional cancer deaths—NASA mission managers would be at risk with everyone else.

But just as Mullane calculated it was "worth it" to him to face a slightly elevated risk of death to enjoy the view, it might not have been the right decision once the welfare of the other six people on board *Atlantis* was factored in.[275] Similarly, although sending RTGs into space or running the RHIC may be "worth it" to the agency personnel engaged in those programs, the level of risk may not be a good bet from the perspective of society as a whole.

The susceptibility of NASA or Brookhaven to moral hazard is likely to be much greater than that of a NASA astronaut in Mullane's position. As sociologist Charles Perrow has noted, "it is hard to have a catastrophe, so the risk to any one set of managers or elites is small, while it is substantial for society as a whole."[276]

What this all means is that agencies need not act irrationally to subject the general public to unacceptable risks of catastrophe. Agencies can make rational decisions that are good for them, even though they may be bad for society over all.[277] As Perrow explains, "[i]n view of the attractions of creating and running risky systems the benefits truly do outweigh the risks for individual calculators."[278]

### *B. Agency Motivation and Divergence from Prescribed Goals*

The possibility of moral hazard faced by *individuals* notwithstanding, one might argue that *agencies* can be expected to make decisions in the public interest so long as they are given the right goals by the legislature or the executive.

---

275. Mullane, perhaps self-servingly, expresses doubt in his autobiography that the bailout system would have worked even if he had been in position to operate it. He writes that its successful operation "presupposed [Commander] Hoot [Gibson] or the autopilot would be able to keep *Atlantis* flying in a straight-ahead, controlled glide. If the vehicle was in a tumble, the G-loads would pin us in the cockpit like bugs on a display board. It wouldn't matter what cockpit—flight deck or mid-deck—we were in." *Id.* at 288–89.

276. CHARLES PERROW, NORMAL ACCIDENTS: LIVING WITH HIGH-RISK TECHNOLOGIES 370 (1984).

277. *Cf.* Dan M. Kahan et. al., *Fear of Democracy: A Cultural Evaluation of Sunstein on Risk*, 119 HARV. L. REV. 1071, 1074–75 (2006) (reviewing CASS R. SUNSTEIN, LAWS OF FEAR: BEYOND THE PRECAUTIONARY PRINCIPLE (2005)) (explaining that under the rational-weigher model, "regulatory intervention is clearly warranted . . . when utility-maximizing individuals are likely to expose others to risks the expected costs of which are not fully borne by those creating them").

278. PERROW, *supra* note 276, at 371. Similarly, in discussing the Space Shuttle *Challenger* disaster, political scientist Kenneth F. Warren has noted, "[i]t is not hard to understand why administrators often take risks that compromise safety, especially when questions of safety involve speculative judgments. The statistical odds of a serious accident happening 'this time' are normally very low, while economic and political realities are usually quite clear." WARREN, *supra* note 258, at 61–62.

Unfortunately, agencies cannot be counted upon to stay true to their charges. Once created, an agency develops its own internal motives. As political scientist Kenneth F. Warren notes, "Every organization has what is referred to as a *natural system,* which consists of basic survival goals that often depart quite radically from the formal system's prescribed goals . . ."[279] Agencies make decisions that are "irrational in terms of the organization's formal objectives yet nevertheless contribute to the organization's ability to cope successfully with its environment."[280] In particular, agencies act in the interests of self-preservation.[281]

As such, it is the nature of agencies to resist political attacks. And agency employees, seeking to protect their reputations and to keep future career prospects open, will fight hard for their agencies.[282] Employees thus fall into line with agency objectives of self-preservation. Because of this, agencies will not necessarily stick to the aims Congress set out for them.

One example of this dynamic in practice comes from the investigation into the loss of NASA's space shuttle *Columbia*. The report of the Columbia Accident Investigation Board emphasized NASA's desire to minimize program delays as an important reason safety was compromised.[283] And NASA wanted to avoid program delays because it saw such delays as politically disastrous, since delays would be fodder for those looking to close down the space shuttle program. NASA's concerns about program delays filtered down to the level of individuals, who understood their own interests would be served by remaining quiet about safety issues that might cause program delays. NASA astronaut Mike Mullane explained, from an astronaut perspective, why women and men may subordinate safety goals and fail to raise issues of risk even when their own lives were on the line:

> We were terrified of saying anything that might jeopardize our place in line to space. We were not like normal men and women who worried about the financial aspects of losing a job, of not being able to make the mortgage payment or pay the kids' tuition. We feared losing a dream, of losing the very thing that made us *us*. When it came to our careers, we were risk averse in the extreme.[284]

---

279. WARREN, *supra* note 258, at 4.

280. *Id.*

281. *Id.* ("Self-preservation is something . . . that all agencies seek, and it is impossible to understand their role in the political system without recognizing this reality.").

282. *Id.* at 39 (citing HERBERT KAUFMAN, ARE GOVERNMENT ORGANIZATIONS IMMORTAL? 7, 11 (1976)).

283. *See* COLUMBIA ACCIDENT INVESTIGATION BD., COLUMBIA ACCIDENT INVESTIGATION BOARD REPORT 200 (2003), http://s3.amazonaws.com/akamai.netstorage/anon.nasa-global/CAIB/CAIB_lowres_full.pdf.

It does not necessarily help one's career to ignore safety issues. Theoretically, if something goes wrong, managers who did not put safety first may be called to answer for their decisions. (Indeed, this was the case following the *Columbia* disaster.)[285] Yet this may not be much of a deterrent. Many dangers may stay latent for years, so that an accident, if one happens, will come about after the manager has moved on.[286] Even without the latency period, elevating safety over other concerns means making trade-offs, and, as Perrow notes, "few managers are punished for not putting safety first even after an accident, but will quickly be punished for not putting . . . agency prestige first."[287]

The prospects of managers being held to answer for their decisions are even more attenuated for risks such as the RHIC and Cassini then they were for the space shuttle program. With Cassini, the global dispersal of plutonium and non-traceability of death would mean that harm suffered would remain an abstraction.[288] And, of course, if the RHIC were to initiate a strangelet conversion of the planet, no one would be around to hold hearings.[289]

### *C. Scientists and Objectivity*

Even if agencies naturally develop motives that diverge from their public charge, and even if actors inside agencies are susceptible to moral hazard, one might argue that where science is involved we need not worry, since scientists employing the scientific method are functionally neutral.

Scientific research is deservedly regarded by our society as a noble pursuit. But the enterprise of modern scientific research has a prosaic side, where it resembles the familiar workaday world of compromise and political wrangling. There is no reason to think, for instance, that particle-accelerator experimentation is immune from the fall-into-line pressure seen at NASA.

Evidence of this kind of influence in accelerator programs can be found in the management philosophy of physicist Samuel Goudsmit, a quantum-physics pioneer and chair of Brookhaven Laboratory's Department of Physics from 1952 to 1960.[290] During his tenure, Brookhaven operated the Cosmotron, the grand particle accelerator of its day.[291] Goudsmit wrote in 1957: "In this new type of work experimental skill must be supplemented by personality traits which enhance loyalty. . . . I feel that we must now deny [the Cosmotron's] use to anyone whose emotional build-up might be detrimental to the cooperative spirit, no mater how good a physicist he is."[292] Given such a code of norms at Brookhaven, a physicist could reasonably expect that coming forth with analysis giving the slightest credence to catastrophic harm scenarios would bring severe career repercussions.

---

284. MULLANE, *supra* note 261, at 214.
285. *See* COLUMBIA ACCIDENT INVESTIGATION BD., supra note 283, at 138–39, 148, 153, 200.
286. PERROW, *supra* note 276, at 370.
287. *Id.*
288. *See supra* Part II.B. (discussing Cassini).
289. *See supra* Part II.C. (discussing the RHIC).
290. *See* Samuel Goudsmit, ARRAY OF CONTEMPORARY AMERICAN PHYSICISTS, https://www.aip.org/history/acap/biographies/bio.jsp?goudsmits (last visited Jan. 24, 2016).

There is a more subtle issue as well. In deciding whether a given activity is safe—or at least worth the risks it imposes—scientists unavoidably make policy judgments. "Safe" is not a number, after all; it's an adjective, and its fittingness is a judgment call. Scientists, in making their judgments, will be subject to the familiar sorts of conflicts of interest and potential for error that can affect agency decision-making generally. What is more, the line between policy and science can be blurry. And as legal scholar Wendy E. Wagner has noted, agencies have demonstrated a predilection to disguise policy choices as science.[293]

Even underneath the level of policy judgments, leading-edge science is more imperfect than it may appear at first glance. While we tend to think of science—and physics in particular—as a matter of hard facts and mathematical certainty, the endeavor can be prone to bad judgment calls like anything else. As physicist Paul Davies has written, "There is a popular misconception that science is an impersonal, dispassionate, and thoroughly objective enterprise. . . . This is, of course, manifest nonsense. Science is a people-driven activity like all human endeavor, and just as subject to fashion and whim."[294]

There is a long history of dogmatic and even wishful thinking that has impinged on scientists' ability to see clearly. A wealth of examples come from the history of nuclear physics. Over the course of the 20th century, elite scientists studying the nucleus repeatedly showed a penchant for construing the natural physical order in such a way that nuclear physics would be safe rather than risky.

In the early 1930s, scientists dismissed the possibility of nuclear fission. When, in 1934, chemist Ida Noddack wrote a paper arguing that the uranium nucleus might be capable of fission, her paper was poorly received.[295] In fact, famed physicist Enrico Fermi dismissed her work as having no possibility of being correct.[296] How Fermi could be so wrong is unclear, but physicist Edward Teller recalled Fermi saying he knew certain things about physics on the basis of "c.i.f.," which stood for *con intuito formidable* meaning "with formidable intuition."[297]

---

291. *See* HARRY COLLINS, GRAVITY'S SHADOW: THE SEARCH FOR GRAVITATIONAL WAVES 551 (2004).

292. *Id.*

293. Wendy E. Wagner, *The Science Charade in Toxic Risk Regulation*, 95 COLUM. L. REV. 1613, 1650 (1995)

294. Paul Davies, *Introduction* to RICHARD P. FEYNMAN, SIX EASY PIECES: ESSENTIALS OF PHYSICS EXPLAINED BY ITS MOST BRILLIANT TEACHER ix (1995).

295. RICHARD RHODES, THE MAKING OF THE ATOMIC BOMB 232 (2012).

Likewise, physicist Otto Frisch considered fission of uranium to be "impossible," and he initially refused to believe the compelling (and correct) arguments made by his aunt, Lise Meitner.[298] Robert Oppenheimer also flatly rejected the possibility of uranium fission, and he offered a number of theoretical reasons why fission could not happen.[299] Oppenheimer only changed his mind when experimentalist Luis W. Alvarez brought him to a laboratory and had him look at an oscilloscope showing signals picked up from a sample of uranium being bombarded with neutrons.[300]

Once the physics community accepted the reality of fission, elite physicists proceeded to shrug off the plausibility of a nuclear chain reaction.[301] Physicist Leo Szilard, the theorist who first proposed the nuclear-chain-reaction concept, recalls how Fermi reacted to his theories: "Fermi thought that the conservative thing was to play down the possibility that [a chain reaction] may happen, and I thought the conservative thing was to assume that it would happen and take all the necessary precautions."[302]

Even once chain reactions were accepted as plausible, safety concerns were brushed aside. As part of the Manhattan Project, Fermi led the experiment to build the world's first artificial nuclear reactor under the stands of the football field at the University of Chicago.[303] Scientists involved in the experiment considered the possibility that the chain reaction would go out-of-control, heat up, and destroy the reactor—what we now term a nuclear meltdown.[304] The prospect was disregarded, however.[305] Physicist Arthur Compton, Fermi's boss on the project, approved the project despite being aware that an out-of-control nuclear reaction in the middle of Chicago would be a disaster.[306] Compton, in fact, took it upon himself *not* to inform the president of the University of Chicago,

296. *Id.*

297. *Id.* at 233.

298. Interview by Charles Weiner with Otto Frisch, Am. Inst. of Physics, New York City, N.Y. (May 3, 1967), *available at* https://www.aip.org/history-programs/niels-bohr-library/oral-histories/4616.

299. RHODES, *supra* note 295, at 274.

300. *Id.*

301. The chain reaction, in which one fissioning nucleus touches off the fission of surrounding nuclei, is the essential mechanism that makes nuclear bombs and nuclear reactors work.

302. RHODES, *supra* note 295, at 281.

303. ARGONNE NAT'L LAB., THOSE EARLY DAYS AS WE REMEMBER THEM: FROM THE METALLURGICAL LABORATORY TO ARGONNE NATIONAL LABORATORY 1 (2013), *available at* http://www.ne.anl.gov/About/early-days/early-days-of-argonne-national-lab.pdf.

304. RHODES, *supra* note 295, at 432–34.

305. *Id.*

306. *Id.* at 433.

legal scholar Robert Maynard Hutchins.[307] According to historian Richard Rhodes, Compton reasoned "that he should not ask a lawyer to judge a matter of nuclear physics"[308]—even as he risked "a small Chernobyl."[309]

Compton later explained himself this way: "Based on [Hutchins's] considerations of the University's welfare[,] the only answer he could have given would have been—No. And this answer would have been wrong. So I assumed the responsibility myself."[310]

Scientists often take the position—implicitly if not explicitly—that scientific research should not be subject to the same sorts of constraints put on other sorts of endeavors. A recurrent example is how scientists regard a claimed need for secrecy because of national security concerns.[311]

In the early stages of World War II, when nuclear physicists were contemplating the potential military applications of nuclear chain reactions, colleagues tried to convince famed physicist Niels Bohr that advances in nuclear physics should be kept secret to prevent Nazi Germany from leveraging them to build a weapon.[312] Bohr, however, was insistent that secrecy not be used in physics.[313] Interestingly, Bohr's argument against secrecy was bolstered by his view of its pointlessness—as physicist Edward Teller remembers, "Bohr insisted that we would never succeed in producing nuclear energy . . . ."[314]

In discussing the question of risk presented by scientific research, Judge Richard A. Posner recounts the example of Barry R. Bloom, dean of Harvard's School of Public Health and a professor of infectious diseases, who criticized journal editors for deciding not to publish the results of biological research that might be instructive for terrorists seeking to create bioweapons.[315] Bloom called such editorial policies "a chilling example of the impact of terrorism on the freedom of inquiry and dissemination of knowledge."[316] Posner, however, observes that Bloom "appears to believe that freedom of scientific research should enjoy absolute priority over every other social value."[317] Posner goes on to note, "Such a belief comes naturally to people who derive career advantages from be-ing able to engage in a particular activity without hindrance, but this belief arbitrarily refuses to weigh costs and so consider the need to make tradeoffs."[318]

---

307. Herbert L. Anderson, *'All In Our Time': Fermi, Szilard and Trinity*, BULL. ATOMIC SCIENTISTS, Oct. 1974, at 40, 43 (Oct. 1974) (quoting ARTHUR COMPTON, ATOMIC QUEST 136 (1956)).

308. RHODES, *supra* note 295, at 432.

309. *Id.* at 433.

310. Anderson, *supra* note 307, at 43 (quoting ARTHUR COMPTON, ATOMIC QUEST 136 (1956)).

311. RHODES, *supra* note 295, at 294.

312. *Id.*

313. *Id.*

314. *Id.*

315. Barry R. Bloom, *Bioterrorism and the University: The Threats to Security and to Openness*, HARV. MAG., Nov.–Dec. 2003, at 48, 51, *quoted in* Posner, *supra* note 154, at 512.

316. *Id.*

317. Posner, *supra* note 154, at 512.

Scientists, swept up in the excitement of the scientific endeavor, may even neglect to take the simplest precautions for their own safety. Some startling examples come from the Trinity Test of the first atomic bomb in the New Mexico desert, where leading Manhattan Project physicists waited for the test 20 miles away from ground zero on Compañia Hill.[319]

"We were told to lie down on the sand," recounted Edward Teller, "turn our faces away from the blast, and bury our heads in our arms. No one complied. We were determined to look the beast in the eye."[320]

Physicist Ernest Lawrence's plan had been to watch the explosion from inside his car, so that at least the windshield would filter the UV rays. Instead, at the last minute, he made the decision to get out of his car—"[E]vidence indeed I was excited!" he reported.[321] "Just as I put my foot on the ground I was enveloped with a warm brilliant yellow white light—from darkness to brilliant sunshine in an instant and as I remember I momentarily was stunned by the surprise."[322]

Nearby, Robert Serber risked blindness by looking straight at the tower. He recalled: "At the instant of the explosion I was looking directly at it, with no eye protection of any kind. . . . The grandeur and magnitude of the phenomenon were completely breathtaking."[323]

Physical chemist George Kistiakowsky was much closer, just five miles from ground zero. Shrugging off official estimates, he figured that the yield of the device would be about 1 kiloton, and he determined he was "very safe" where he was.[324] Taking a pass on the concrete dugout, Kistiakowsky stood on an earthen mound to watch the detonation.[325] Unfortunately for Kistiakowsky, he was very wrong about the yield. The explosion measured 18.6 kilotons.[326] How Kistiakowsky fared is unclear.

---

318. *Id.*

319. Larry Calloway, *The Nuclear Age's Blinding Dawn*, ALBUQUERQUE J., July 1995, at 8, http://www.abqjournal.com/trinity/trinity1.pdf (location of Compañia Hill).

320. RHODES, *supra* note 295, at 666.

321. E.O. Lawrence, Thoughts by E.O. Lawrence (July 16, 1945), *in* CENTER FOR THE STUDY OF INTELLIGENCE, CIA, CSI 98-10001, THE FINAL MONTHS OF THE WAR WITH JAPAN: SIGNALS INTELLIGENCE, U.S. INVASION PLANNING, AND THE A-BOMB DECISION app. C, Document 12, 1369 (1998).

322. *Id.*

323. RHODES, *supra* note 295, at 673.

324. *Id.* at 669.

325. *Id.* at 667.

326. ROBERT SERBER, THE LOS ALAMOS PRIMER: THE FIRST LECTURES ON HOW TO BUILD AN ATOMIC BOMB 60 (1992).

There seems to be no recorded account of whether he suffered any immediate injuries.[327]

Scientists on the cutting edge of knowledge can be wrong. They can make incorrect assumptions. They can take unnecessary chances. They can over-estimate the certainty of their conclusions and underestimate the risks of their activities. In sum, the supposition that science involves neutral decision-making cannot ameliorate concerns above catastrophic risks posed by leading-edge experiments.

### *D. Behavioral Economics and Psychology*

Up to this point, I have discussed the corruptibility of agency scientists and managers in making decisions that set their own self interests and passions against the interests of the public at large and, in some cases, their own common sense. In this section, I want to discuss how common sense itself can become warped, such that decision-makers, despite diligence and good intentions, can inadvertently drift toward bad choices. In so doing, I will discuss how empirically validated insights from behavioral economics and cognitive psychology show how people fall into predictable patterns of thought that lead to irrational judgments.

The first observation to make is that, as a general matter, people are irrational when it comes to judging risk.[328] This is a familiar concept in the regulation literature, as people's irrationality with risk is one reason to expect that free markets will not reach efficient results. Because of this, irrationality in dealing with risk may be regarded as a compelling rationale for giving agencies more power—allowing them to step in with regulations to trump free-market transactions. But irrationality is not the exclusive province of consumers. Agency managers and scientists can be irrational in evaluating risk as well. A survey of behavioral economic concepts, in fact, shows that they have particular relevance for the kinds of catastrophic and irreversible ultrahazards that can be posed by government science experiments.

One pattern of irrationality that people tend to exhibit when thinking about loss is "myopia bias"—an undue focus on the here and now and a corresponding lack of concern for the future.[329] The myopia bias has been documented by a substantial body of experimental work, find-ing, as an empirical matter, that "people value the avoidance of immediate or nearly immediate losses far more strongly than the avoidance of losses even in the not-too-distant future."[330] Legal scholar David A. Dana suggests that this has ramifications for environmental policy. He writes that "decisionmakers will weigh immediate economic losses more heavily than they should in comparison with non-immediate health and environmental losses."[331] By the same token, myopia bias suggests that agency decision-makers—despite a genuine desire to make judgments in the public interest—may give much more weight to the prospect of the immediate loss of a government program than to the longer-term problem of additional cancers or a latent threat to the planet.

---

327. Kistiakowsky lived long afterward, dying at age 82 after a battle with cancer. *See G.B. Kistiakowsky is Dead at 82; Bomb Pioneer Sought Nuclear Ban*, N.Y. TIMES, Dec. 8, 1982, http://www.nytimes.com/1982/12/08/obituaries/gb-kistiakowsky-is-dead-at-82-bomb-pioneer-sought-nuclear-ban.html. Although radiation can cause cancer, it would be impossible to tell if Kistiakowsky's disease was related to his exposure at Trinity. Considering his lifetime spent working with hazardous chemicals, it seems likely that any number of occupational exposures, in addition to natural causes, could have been the culprit.

328. *See, e.g.*, CASS R. SUNSTEIN, RISK AND REASON: SAFETY, LAW, AND THE ENVIRONMENT 28–52 (2002) [hereinafter SUNSTEIN, RISK] (discussing in detail irrationality in valuing risk).

329. *See, e.g.*, Dana, *supra* note 26, at 1324–25.

Another pattern in people's flawed perception of risk is what is known as "probability neglect," by which people will be overly concerned with sure losses and less concerned with unsure losses, even if those unsure losses are of a much greater magnitude.[332] The relevance to the Cassini and RHIC cases is clear: Both involve uncertain and apparently low probabilities of coming to fruition, but both raise the specter of great harm.[333] Thus, we should suspect that decision-makers may underweight their potential for disaster.

Related to and partially overlapping with probability neglect is the effect known as "optimism bias," "whereby people may believe, even in the absence of any factual basis, that with time they will find a costless means to avoid future risks."[334] This suggests that decision-makers may be untroubled—or at least *under*troubled—by the prospect of creating the latent potential for harm when investing billions of dollars in large-scale experimental programs with unresolved risk questions.

Another failure of rationality is what is known as the "availability bias" or "availability heuristic." This effect causes people to evaluate the likelihood of something happening on the basis of "the ease with which instances or occurrences can be brought to mind."[335] Indeed, it is difficult to come up with an example of something more unprecedented, more *unavailable* in a cognitive sense, than a strange-matter disaster. The possibility of our planet being crushed into a ball that could fit under the St. Louis Arch is not just bad.[336] It is bad-movie bad. The mental picture is so absurd, it is hard to take seriously. Philosopher Nick Bostrom has said about existential risks: "We might find it hard to take them as seriously as we should simply because we have never yet witnessed such disasters. Our collective fear-response is likely ill calibrated to the magnitude of threat[.]"[337]

---

330. *Id.*

331. *Id.* at 1325.

332. Kahan et al., *supra* note 277, at 1077–78.

333. *See supra* Part II.B–C (regarding Cassini and the RHIC).

334. Dana, *supra* note 26, at 1325.

335. Amos Tversky & Daniel Kahneman, *Judgment Under Uncertainty: Heuristics and Biases*, 185 SCIENCE 1124, 1127 (1974); Molly J. Walker Wilson, *Cultural Understandings of Risk and the Tyranny of the Experts*, 90 OR. L. REV. 113, 134 (2011).

336. St. Louis's Gateway Arch is 192 meters tall and wide. *See* William V. Thayer, *The Mathematics and Architecture of the Saint Louis Arch*, http://www.jug.net/wt/arch.htm (last visited Jan. 24, 2016). By comparison, the remaining diameter of Earth after a strange-matter conversion has been estimated at 100 meters. *See* REES, *supra* note 79, at 121.

Put together, these behavioral and cognitive phenomena describe a tendency for people to irrationally neglect uncertain and unlikely risks, particularly those for which there is no precedent—in other words, exactly the kinds of hypothetical laboratory catastrophes that might be at issue with federal agency science research.

A plausible objection to this argument is that, despite the empirical evidence for such psychological effects among people evaluating risk, we have no reason to think that these kinds of effects will prevail in the government agency context. Specifically, one might suppose that irrational evaluations done by individuals will not necessarily translate to irrational evaluation by groups of people. And it is how people behave in groups that is relevant to the agency context. Research, however, indicates "the absence of any clear-cut difference between individual and group decisionmaking with respect to the influence of cognitive biases."[338] Moreover, there is the possibility—which seems just as plausible—that the group context *enhances* these cognitive biases.[339]

Another potential objection is that these biases do not apply to experts in the same way that they do to the rest of the population; that is, with their advanced training and knowledge, agency managers and scientists, as experts, are not vulnerable to cognitive biases in the way that everyone else is. This is the general position, for instance, taken by Cass Sunstein.[340] But the idea that experts can rise above their own cognitive humanness seems dubious. Molly J. Walker Wilson writes, "Because experts are not outside of the cultural milieu, they, like others, are influenced by inevitable cognitive patterns, emotion-based responses, and political and moral concerns."[341] In fact, there is substantial evidence that experts themselves, when calculating probabilities, use heuristics to deal with the inevitable uncertainties that come with a dataset.[342]

---

337. Nick Bostrom, *Existential Risks: Analyzing Human Extinction Scenarios and Related Hazards*, 9 J. EVOLUTION & TECH. 2 (2002) (footnote omitted), *available at* http://jetpress.org/volume9/risks.html.

338. Dana, *supra* note 26, at 1330 n.42 (citing Norbert L. Kerr et al., *Bias in Judgment: Comparing Individuals and Groups*, 103 PSYCHOL. REV. 687, 697 (1996)).

339. *Id.* at 1330.

340. *See* Kahan et. al, *supra* note 277, at 1093 (summarizing Sunstein's critique); Dana, *supra* note 26, at 1330 n.42 (same). Note, however, that Sunstein does acknowledge that experts can be induced to self-censor when it comes to risk issues. *See* SUNSTEIN, RISK, *supra* note 328, at 87–88 (pointing, as examples, to a medical researcher reluctant to talk about Lyme disease diagnoses who told the *New York Times*, "If you quote me saying these things, I'm as good as dead," and a sociologist skeptical of accepted wisdom about bovine spongiform encephalitis, who said about his interactions, "You get made to feel like a pedophile").

341. Wilson, *supra* note 335, at 188.

Dana points out three reasons why experts should not be viewed as exempt from cognitive biases. First, he notes, empirical studies indicate that cognitive biases are hard-wired into the brain, and thus experts will be unable to completely shake these biases.[343] Although experts may be less vulnerable than laypersons with regard to some biases, the differences are small and, at least with regard to loss aversion and myopia bias, no studies indicate experts fare any better.[344] Second, Dana notes that cognitive biases have the most substantial effect in the context of very complex questions, and these are exactly the kinds of questions that experts in government agencies deal with: "Their reaction—their very human reaction—to such complexity is to resort to cognitive shortcuts as a means to bring order (if not accurate analysis) to confusing masses of information and critical information gaps."[345] Third, Dana makes the point that experts tend to suffer the effects of one particular bias more than non-experts: "overconfidence in their own judgments."[346] This in turn "may translate into greater distortions in their decisionmaking than in the decisionmaking of non-experts."[347]

There is also the influence of culture. An anthropological study of high-energy physicists found that physicists tend to believe there are no cultural influences on their scientific work.[348] But research has shown that experts do come to conclusions about risk that are explainable only in terms of culture. Where experts disagree with one another about risks, their views sort according to variables such as political ideology, institutional affiliation, and gender.[349]

Sociologist and public-affairs scholar Diane Vaughan, who made an in-depth study on the causes of the *Challenger* disaster, has explained that people make risk assessments "through the filtering lens of individual worldview."[350] And the worldview of experts can lead to irrational results because experts, in spite of themselves, will labor to preserve their worldview against information that contradicts it.[351] Another scholar who

342. Donald T. Hornstein, *Reclaiming Environmental Law: A Normative Critique of Comparative Risk Analysis*, 92 COLUM. L. REV. 562, 610–11 (1992).

343. Dana, *supra* note 26, at 1332.

344. *Id.*

345. *Id.* at 1332–33.

346. *Id.* at 1333.

347. *Id.*

348. SHARON TRAWEEK, BEAMTIMES AND LIFETIMES 78 (1992).

349. Kahan et al., *supra* note 277, at 1093.

350. DIANE VAUGHAN, THE CHALLENGER LAUNCH DECISION 62 (1996).

351. *Id.* at 63 ("They may puzzle over contradictory evidence but usually succeed in pushing it aside—until they come across a piece of evidence too fascinating to ignore, too clear to misperceive, too painful to deny, which makes vivid still other signals they do not want to see, forcing them to alter and surrender the worldview they have so meticulously constructed.") (citations omitted).

investigated the *Challenger* disaster, Charles Perrow, has written along similar lines: "Managers come to believe their own rhetoric about safety first because information indicating otherwise is suppressed for reasons of organizational politics."[352]

Applying these insights to cases such as Cassini and the RHIC is straightforward. We should expect that the employees of these agencies will view their projects as propitious, their aims as noble, and their preparatory work as diligent. Under such circumstances, it would not be unexpected for agency decision-makers to miss signs pointing to the existence of real danger.

### *E. Reasons to Believe the Political Process Will Not Ameliorate Inappropriate Judgments of Agencies*

Notwithstanding the foregoing argument that agencies and scientists can get catastrophic risk decisions wrong, one might argue that any deficiencies in agency decision-making are ameliorated by the ultimate control that the President and Congress have over administrative agencies. Thus, it could be argued, the political process can compensate for problems such as agency self-interest and the cognitive biases of agency decision makers.

Unfortunately, there are many reasons to believe that such confidence in the political system is misplaced.

At the outset, it must be noted that citizens and actors in the political system are subject to the same cognitive biases as experts, with the usual account being that non-experts are more irrational, not less.[353] The greater irrationality of non-experts can be attributed to a general illusion of safety, which is driven by lack of relevant experience. As Sunstein has noted, people are naturally drawn to having an illusion of safety when it comes to low-probability risks.[354] A research team conducting a behavioral experiment made a similar observation, stating that "there seems to be something in human cognition that sometimes treats tiny probabilities, even of extreme disaster, as worth less than their expected value."[355]

Non-experts' lack of experience with catastrophic and irreversible risks leads to the downplaying of risks. Non-experts tend to evaluate risk

352. PERROW, *supra* note 276, at 370.

353. Kahan et al., *supra* note 277, at 1079.

354. CASS R. SUNSTEIN, WORST-CASE SCENARIOS 22 (2007) [*hereinafter* SUNSTEIN, WORST-CASE]; *see also* Sunstein, *Irreversible and Catastrophic*, *supra* note 26, at 16 (citing Gary H. McClelland, William D. Schulze, and Don L. Coursey, *Insurance for Low-Probability Hazards: A Bimodal Response to Unlikely Events*, 7 J. RISK & UNCERTAINTY 95, 95, 102 (1993) ("[T]here seems to be something in human cognition that sometimes treats tiny probabilities, even of extreme disaster, as worth less than their expected value.").

355. Sunstein, *Irreversible and Catastrophic*, *supra* note 26, at 16 (citing Gary H. McClelland et al., *Insurance for Low-Probability Hazards: A Bimodal Response to Unlikely Events*, 7 J. RISK & UNCERTAINTY 95, 95, 102 (1993).

on the basis of intuition, and intuitions about risk are highly unreliable,[356] tend to be rapid, and rest on personal experience.[357] Without relevant experience, which all non-experts lack, we may believe an unlikely event is not worthy of our attention.[358] This can lead to serious judgment errors.[359] Expecting citizens, who naturally succumb to these analytical shortcuts, to put political pressure on an agency, which itself has made an irrational decision on risk, is asking too much. And since politicians are motivated to act primarily by public engagement,[360] relying on the political process to correct erroneous agency decisions on risk is untenable.

Citizen non-engagement is not the only factor likely to render political oversight inadequate. The harms associated with the disaster scenarios considered here have long latency times. Years may go by before injury comes to fruition. Politicians are therefore underincentivized to consider exotic risks in the present because it is highly unlikely they will face scrutiny in the future for ignoring those risks. As Posner writes, "Politicians with limited terms of office, and thus foreshortened political horizons, are likely to discount low risk disaster possibilities steeply since the risk of harm to their careers from failing to take precautionary measures is truncated."[361]

As has often been observed, it can take a crisis to get people to respond appropriately to a threat.[362] But that is little consolation for large-scale irreversible harms. The RHIC strange-matter scenario, of course, represents an extreme in this respect.

## V. COURTS AS THE SOLUTION

In this Article, I have set out a detailed case for why it is problematic to exempt from judicial review decisions in which an agency decides for itself whether its own experimental program presents an acceptable risk of catastrophe. With the problem identified, the solution is straightforward: The legal system must ensure meaningful judicial review of such risks.

Courts are an excellent venue for addressing small-probability/large-harm risks. Why? The judiciary provides a unique opportunity to get beyond mistake-inducing mental heuristics. Courts lack the personal and institutional biases to be found inside agencies. And unlike

356. *See* SUNSTEIN, RISK, *supra* note 328, at 29.
357. *See* SUNSTEIN, WORST-CASE, *supra* note 354, at 5.
358. *See id.* at 6.
359. *See id.* at 6–7; *see also* SUNSTEIN, RISK, *supra* note 328, at 38 (discussing the irrational minimization of risks).
360. Or the anticipation of public engagement in positive or negative ways, as may be brought on by special interests and mediated through the use of campaign funds.
361. Posner, *supra* note 154, at 514.
362. *See, e.g.*, SUNSTEIN, WORST-CASE, *supra* note 354, at 58.

the public, whose economy of attention is necessarily limited, courts have the luxury of being able to explore a problem in depth. Moreover, civil discovery and the adversarial process can uncover important facts that agency insiders may be prone to ignore.[363] And the questions and arguments of opposing counsel can sharpen the factual picture. Simply put, courts are good for providing review of agency experiment-risk questions for the same reasons courts are good for contract disputes, criminal charges, and everyday negligence claims: Courts have the time, resources, and independence needed to get to a fair result.

Moreover, courts should not be deterred from providing judicial review out of a concern that doing so is difficult. Courts need not understand the science as scientists do to gauge the reliability of the scientists' risk-assessment work in the aggregate.[364] For instance, no scientific expertise is needed for a generalist judge to determine whether risk-assessment work carries hallmarks of self-interest or hastiness.

Ensuring the availability of judicial review is not problematic as a technical matter. Both Congress and the courts have ready means.

Congress could explicitly invite judicial review either by amending the APA or the FTCA to guarantee reviewability of non-rulemaking, non-adjudicatory agency decisions, regardless of their discretionary nature, where the proposed course of agency action allegedly constitutes a catastrophic and irreversible ultrahazard.[365] Such a distinct authorization for judicial review—by being limited to threats that are catastrophic, irreversible, and ultrahazardous—would not threaten the efficient operation of the administrative state as a whole by subjecting every discretionary bureaucratic decision to litigation. Moreover, such a judicial-review authorization is reasonable because, by its own terms, it is limited to circumstances of surpassing importance.

Although amending the APA or FTCA would fix the problem completely, congressional action is not necessary. Nor would one assume it to be likely, absent a catastrophe of sufficient magnitude to set the political process in motion. I argued above that political oversight is unlike-ly to provide a sufficient check on low-probability/high-harm risks of agency conduct.[366] For the same reason, it seems unpropitious to look to congressional action to dragoon the courts into providing judicial review in cases of catastrophic/irreversible agency hazards.

---

363. As a helpful point of comparison, it is instructive to consider the substantial weaknesses in safety analyses that the Busza and Dar groups pointed out in each others' papers, even though the Busza and Dar groups were not adversarial, as both groups argued that the RHIC was safe. *See supra* note 122 and accompanying text.

364. In previous work, I have described in detail how courts can providing fair and meaningful judicial review of science-intensive risk assessment where the subject-matter is recondite to the generalist judge. *See* Johnson, *supra* note 94, at 885 ("Using a kind of meta-analysis, courts should gauge the risk that scientific judgments are wrong. Relevant subjects of inquiry include organizational culture, group politics, and psychological context. The particular aspects of scientific arguments should also be scrutinized on a meta level. Relevant issues here include the newness of underlying theory, the complexity of the chain of argument, the likely reliability of underlying data, and so on. Also relevant is what history has to say about the durability of pronouncements made in the field.").

365. For discussion of what constitutes a "catastrophic and irreversible ultrahazard," see *supra* Part II.A.

The institution best-suited to assert the need for judicial review is the judiciary itself. Even if statutory law does not explicitly endorse judicial review of non-adjudicatory/non-regulatory agency decisions, statutory law clearly does not prohibit such review either. That full measure of foreclosure must come, if it comes at all, from other court-created doctrines—such as, in particular, sovereign immunity or political-question doctrine.[367] As court-crafted doctrines, they can and should be limited—or even reworked—where prudence and logic requires.[368] Moreover, any interpretation of the APA that would preclude judicial review can easily be undone. The APA's provisions on judicial review are general enough that a court can and should adjust its interpretation of the statute where, as here, the sorts of agency decisions under scrutiny are of a type not contemplated by the APA's framers.

As a final matter in discussing solutions, it should be noted that there is no substantive barrier to a court having the authority to issue an injunction against conduct that threatens catastrophic harm. That is something the courts already have. With today's burgeoning administrative state, we are used to seeing agencies play the role of ex-ante accident preventers. But courts have long had the inherent equitable authority to enjoin negligence. An excellent example is *Harris Stanley Coal & Land Co. v. Chesapeake & Ohio Railway Co.*,[369] a case handed down in 1946—the same year the FTCA and APA were passed.[370] The dispute arose from the Chesapeake & Ohio's operation of railroad tracks above Harris Stanley Coal's underground mine. The railroad wanted an injunction to prohibit the mine from "pulling the pillars"—a technique in which columns of coal originally left intact to support the mine's ceiling are demolished to get the additional coal.[371] The railroad alleged that pillar-pulling could cause the earth under the rail lines to subside, leading to a derailment.[372] The court agreed and enjoined the mine. In doing so, the court exemplified a judicial capacity for prudence in looking at low-probability/high-harm risk, observing, "It may be that such disaster could

366. *See supra* Part IV.E.
367. *See supra* Parts III.B.1, III.B.5.
368. Arguments can be made that other aspects of the law might preclude judicial review, such as standing or the federal-contractor doctrine. See *supra* notes 181–183 and accompanying text listing some such aspects. Here, too, case law should be interpreted and doctrine evolved in accordance with need and logic.
369. 154 F.2d 450 (6th Cir. 1946).
370. *See supra* notes 10, 13.
371. *See Harris Stanley Coal & Land Co.*, 154 F.2d at 452.
372. *Id.*

occur only upon a concatenation of circumstances of not too great probability, and that the odds are against it. It is common experience, however, that catastrophies occur at unexpected times and in unforeseen places."[373]

## VI. CONCLUSION

Stewardship of administrative law has largely concerned two functions that agencies perform—rulemaking and adjudication. Yet many agency activities fall outside these two functions. And some of these activities may entail ultrahazardous risks of irreversible catastrophe. In this Article, I have looked at agency science-research activity that plausibly creates such risks. In so doing, I have sought to highlight the need for effective legal constraints on these activities. That need arises because when it comes to low-probability/high-harm scenarios occasioned by an agency's own conduct, that agency is unlikely to adequately safeguard the public interest.

As case studies, I provided an in-depth look at two agency-science programs said to run a small chance of causing enormous catastrophe: dispersal of large quantities of plutonium-238 from a NASA space probe and destruction of the Earth by a DOE particle collider. Despite their seeming exoticness, there is no reason to believe these examples are idiosyncratic. The CDC, for instance, recently owned up to serious incidents of mishandling pathogens such as H1N1 influenza and smallpox, either of which has the potential to cause a pandemic if released to the outside world.[374] Moreover, these sorts of ultrahazardous-risk issues are unlikely to go away on their own. To the contrary, we should expect them to proliferate.

The more successful science is in gaining knowledge of our natural world by means of the experimental tools we have at our disposal, the more ambitious science must be to reach the next plateau of knowledge. Each experimental program tends to be bigger, more expensive, and more complex than the last. And the more expensive science becomes, the greater the need for taxpayer-funded government agencies to undertake it. Thus, a refusal of the law to deal with agency-created risk becomes increasingly undesirable.

In closing, I should note that science-experiment risk is not the only species of non-adjudicatory/non-regulatory agency activity that implicates low-probability/high-harm risks. Disaster can loom, for instance,

---

373. *Id.* at 453–54 (internal citations omitted).

374. *See, e.g.*, Carlos Moreno, Op-ed., *CDC Mishaps Show Live Flu Viruses Are Nothing to Play With*, REUTERS, Jul 28, 2014, http://blogs.reuters.com/great-debate/2014/07/28/want-to-avoid-a-pandemic-heres-a-good-way-to-start/ (discussing recent revelations of mishandling live pathogens).

where agencies build levies and dams[375] or intentionally ignite fires for land-management purposes.[376] I hope the analysis in this Article will be helpful to thinking about catastrophic risks from these and other non-governing agency actions.

The courts have long been the ready bulwark to protect the public interest from government overreach. They should not shy from this responsibility when it comes to non-adjudicatory/non-regulatory agency action—particularly where an agency may pose a small risk of great disaster. What particular precautions are needed in any given case will be a difficult question. There is, however, an easily discerned meta-precaution: We can ensure that the courts engage with these cases and bring to bear on them the same procedures for entertaining claims, resolving disputed facts, and providing relief that the courts have long provided in less outré contexts.

375. An example is the failure of the New Orleans levee system, built by the U.S. Army Corps of Engineers, in Hurricane Katrina. *See, e.g.*, Campbell Robertson & John Schwartz, *Decade After Katrina, Pointing Finger More Firmly at Army Corps*, N.Y. TIMES, May 23, 2015, http://www.nytimes.com/2015/05/24/us/decade-after-katrina-pointing-finger-more-firmly-at-army-corps.html.

376. An example is an out-of-control wildfire started by the Department of the Interior's Bandelier National Monument. *See Bandelier Conducts First Prescribed Fire Since 2000*, NAT'L PARK SERV., http://www.nps.gov/fire/wildland-fire/connect/fire-stories/2006-parks/bandelier-national-monument.cfm ("In May of 2000, a planned prescribed fire ignited by the Bandelier fire staff escaped containment and became the Cerro Grande Wildland Fire. This 40,000+ acre fire destroyed over 240 homes in nearby Los Alamos and damaged public and private lands on an unprecedented scale.") (last visited Jan. 24, 2016).